\documentclass[a4paper,twocolumn,11pt]{quantumarticle}
\pdfoutput=1

\usepackage{amsmath}
\usepackage{hyperref}
\usepackage{physics}
\usepackage{graphicx}
\usepackage{float}
\usepackage{dcolumn}
\usepackage{subfigure}
\usepackage{dsfont}
\usepackage[english]{babel}
\usepackage[utf8]{inputenc}
\usepackage[T1]{fontenc}
\usepackage{mathrsfs}
\usepackage{amsfonts}
\usepackage{bbm}
\usepackage{bm}
\usepackage{enumerate}
\usepackage[dvipsnames]{xcolor}
\usepackage{subdepth}
\usepackage{fnpct}
\usepackage{booktabs}
\usepackage{multirow}
\usepackage[math]{cellspace}
\usepackage[numbers]{natbib}
\usepackage{tikz}

\newcommand{\ketu}{%
\begin{tikzpicture}[baseline={([yshift=-2.5pt]current bounding box.center)},scale=0.025,every node/.style={scale=0.1}, every path/.style={line width=0.2mm}]
\draw    (0,60) -- (5,60) ;
\draw    (0,40) -- (5,40) ;
\draw    (0,20) -- (5,20) ;
\draw    (0,0) -- (5,0) ;
\draw    (4.90,-0.07) .. controls (15,3) and (15,17) .. (4.90,20) ;
\draw    (4.90,39.93) .. controls (15,43) and (15,57) .. (4.90,60) ;
\end{tikzpicture}}%

\newcommand{\brau}{\begin{tikzpicture}[baseline={([yshift=-2.5pt]current bounding box.center)},scale=0.025,every node/.style={scale=0.1}, every path/.style={line width=0.2mm}]
\draw    (10,60) -- (15,60) ;
\draw    (10,40) -- (15,40) ;
\draw    (10,20) -- (15,20) ;
\draw    (10,0) -- (15,0) ;
\draw    (10.10,-0.07) .. controls (0,3) and (0,17) .. (10.10,20) ;

\draw    (10.10,39.93) .. controls (0,43) and (0,57) .. (10.10,60) ;
\end{tikzpicture}}%

\newcommand{\ketv}{\begin{tikzpicture}[baseline={([yshift=-2.5pt]current bounding box.center)},scale=0.025,every node/.style={scale=0.1}, every path/.style={line width=0.2mm}]
\draw    (0,60) -- (5,60) ;
\draw    (0,40) -- (5,40) ;
\draw    (0,20) -- (5,20) ;
\draw    (0,0) -- (5,0) ;
\draw    (4.90,-0.07) .. controls (15,3) and (15,57) .. (4.90,60) ;
\draw    (4.90,19.93) .. controls (10,23) and (10,37) .. (4.90,40) ;
\end{tikzpicture}}%

\newcommand{\brav}{\begin{tikzpicture}[baseline={([yshift=-2.5pt]current bounding box.center)},scale=0.025,every node/.style={scale=0.1}, every path/.style={line width=0.2mm}]
\draw    (10,60) -- (15,60) ;
\draw    (10,40) -- (15,40) ;
\draw    (10,20) -- (15,20) ;
\draw    (10,0) -- (15,0) ;
\draw    (10.10,-0.07) .. controls (0,3) and (0,57) .. (10.10,60) ;
\draw    (10.10,19.93) .. controls (5,23) and (5,37) .. (10.10,40) ;
\end{tikzpicture}}%
\newcommand{\braz}{\begin{tikzpicture}[baseline={([yshift=-2.5pt]current bounding box.center)},scale=0.025,every node/.style={scale=0.1}, every path/.style={line width=0.2mm}]
\draw    (10,60) -- (15,60) ;
\draw    (10,40) -- (15,40) ;
\draw    (10,20) -- (15,20) ;
\draw    (10,0) -- (15,0) ;
\draw    (10.10,-0.07) .. controls (0,3) and (0,17) .. (10.10,20) ;
\draw    (10.10,19.93) .. controls (0,23) and (0,37) .. (10.10,40) ;
\draw    (10.10,39.93) .. controls (0,43) and (0,57) .. (10.10,60) ;
\end{tikzpicture}}%

\newcommand{\ketz}{\begin{tikzpicture}[baseline={([yshift=-2.5pt]current bounding box.center)},scale=0.025,every node/.style={scale=0.1}, every path/.style={line width=0.2mm}]
\draw    (0,60) -- (5,60) ;
\draw    (0,40) -- (5,40) ;
\draw    (0,20) -- (5,20) ;
\draw    (0,0) -- (5,0) ;
\draw    (4.90,-0.07) .. controls (15,3) and (15,17) .. (4.90,20) ;
\draw    (4.90,19.93) .. controls (15,23) and (15,37) .. (4.90,40) ;
\draw    (4.90,39.93) .. controls (15,43) and (15,57) .. (4.90,60) ;
\end{tikzpicture}}%

\newcommand{\boxfourlegs}[1]{\begin{tikzpicture}[baseline={([yshift=-4.3pt]current bounding box.center)},scale=0.025,every node/.style={scale=0.9}, every path/.style={line width=0.2mm}]

\draw    (0,20) -- (50,20) ;
\draw    (0,0) -- (50,0) ;

\draw[rounded corners=1] (10,35) rectangle (40,65);

\draw    (0,60) -- (10,60) ;
\draw    (0,40) -- (10,40) ;

\draw    (40,60) -- (50,60) ;
\draw    (40,40) -- (50,40) ;
\node at (25,50) {\scriptsize $#1$};
\end{tikzpicture}}%

\newcommand{\boxtwotrA}[1]{\begin{tikzpicture}[baseline={([yshift=-2.5pt]current bounding box.center)},scale=0.025,every node/.style={scale=0.9}, every path/.style={line width=0.2mm}]
\draw[rounded corners=1] (10,35) rectangle (40,65);

\draw    (5,60) -- (10,60) ;
\draw    (5,40) -- (10,40) ;

\draw    (40,60) -- (45,60) ;
\draw    (40,40) -- (45,40) ;

\draw    (5.09,39.93) .. controls (0,43) and (0,57) .. (5.09,60) ;
\draw    (44.91,39.93) .. controls (50,43) and (50,57) .. (44.91,60) ;

\node at (25,50) {\scriptsize $#1$};
\end{tikzpicture}}%

\newcommand{\boxtwotrB}[1]{\begin{tikzpicture}[baseline={([yshift=-2.5pt]current bounding box.center)},scale=0.025,every node/.style={scale=0.9}, every path/.style={line width=0.2mm}]
\draw[rounded corners=1] (10,35) rectangle (40,65);

\draw    (5,60) -- (10,60) ;
\draw    (5,40) -- (10,40) ;

\draw    (40,60) -- (45,60) ;
\draw    (40,40) -- (45,40) ;

\draw    (5.1,60) .. controls (0,80) and (50,80) .. (44.91,60) ;
\draw    (5.1,39.93) .. controls (0,20) and (50,20) .. (44.91,40) ;

\node at (25,50) {\scriptsize $#1$};
\end{tikzpicture}}%

\newcommand{\ketub}{\begin{tikzpicture}[baseline={([yshift=-2.5pt]current bounding box.center)},scale=0.025,every node/.style={scale=0.1}, every path/.style={line width=0.2mm}]
\draw    (0,60) -- (5,60) ;
\draw    (0,40) -- (5,40) ;

\draw    (4.90,39.93) .. controls (15,43) and (15,57) .. (4.90,60) ;
\end{tikzpicture}}

\newcommand{\braub}{\begin{tikzpicture}[baseline={([yshift=-2.5pt]current bounding box.center)},scale=0.025,every node/.style={scale=0.1}, every path/.style={line width=0.2mm}]
\draw    (10,60) -- (15,60) ;
\draw    (10,40) -- (15,40) ;

\draw    (10.10,39.93) .. controls (0,43) and (0,57) .. (10.10,60) ;
\end{tikzpicture}}%

\newcommand{\idtwo}{\begin{tikzpicture}[baseline={([yshift=-2.5pt]current bounding box.center)},scale=0.025,every node/.style={scale=0.1}, every path/.style={line width=0.2mm}]
\draw    (0,60) -- (15,60) ;
\draw    (0,40) -- (15,40) ;

\end{tikzpicture}}%

\newcommand{\boxonetr}[1]{\begin{tikzpicture}[baseline={([yshift=-2.5pt]current bounding box.center)},scale=0.025,every node/.style={scale=0.9}, every path/.style={line width=0.2mm}]
\draw[rounded corners=1] (10,20) rectangle (30,40);

\draw    (0,30) -- (10,30) ;
\draw    (30,30) -- (40,30) ;
\draw    (0.1,30) .. controls (10,5) and (30,5) .. (39.91,30) ;

\node at (20,30) {\scriptsize $#1$};
\end{tikzpicture}}%

\newcommand{\fourops}[4]{%
\begin{tikzpicture}[baseline={([yshift=-2.5pt]current bounding box.center)},
    scale=0.025,
    every node/.style={scale=0.6},
    every path/.style={line width=0.2mm}
]
  \draw (0,60) -- (5,60);
  \draw (0,40) -- (5,40);
  \draw (0,20) -- (5,20);
  \draw (0,0)  -- (5,0);

  \node[right] (n1) at (5,60) {$#1$};
  \node[right] (n2) at (5,40) {$#2$};
  \node[right] (n3) at (5,20) {$#3$};
  \node[right] (n4) at (5,0)  {$#4$};

  \draw (n1.east) ++(0.5,0) -- ++(5,0);
  \draw (n2.east) ++(0.5,0) -- ++(5,0);
  \draw (n3.east) ++(0.5,0) -- ++(5,0);
  \draw (n4.east) ++(0.5,0) -- ++(5,0);
\end{tikzpicture}}

\newcommand{\leftparenthesis}{%
\begin{tikzpicture}[scale=0.3, baseline={([yshift=-2.5pt]current bounding box.center)}]
  \draw[line width=0.3pt] 
    (1,6) .. controls (0,6) and (0,5) .. (0,3)
           .. controls (0,1) and (0,0) .. (1,0);
\end{tikzpicture}%
}

\newcommand{\rightparenthesis}{%
\begin{tikzpicture}[scale=0.3, baseline={([yshift=-2.5pt]current bounding box.center)}]
  \draw[line width=0.3pt] 
    (0,6) .. controls (1,6) and (1,5) .. (1,3)
           .. controls (1,1) and (1,0) .. (0,0);
\end{tikzpicture}%
}

\newcommand{\brauidtimesfour}[4]%
{\begin{tikzpicture}[baseline={([yshift=-2.5pt]current bounding box.center)},scale=0.025,every node/.style={scale=0.6}, every path/.style={line width=0.2mm}]
  \draw    (0,60) -- (5,60) ;
\draw    (0,40) -- (5,40) ;
\draw    (0,20) -- (5,20) ;
\draw    (0,0) -- (5,0) ;

\node[right]  at (5,60) {$#1$};
\node[right]  at (5,40) {$#2$};
\node[right]  at (5,20) {$#3$};
\node[right]  at (5,0) {$#4$};

\draw    (18,60) -- (23,60) ;
\draw    (18,40) -- (23,40) ;
\draw    (18,20) -- (23,20) ;
\draw    (18,0) -- (23,0) ;

\draw (0,60) .. controls (5,45) and (25,50) .. (23,40); 
\end{tikzpicture}}%

\newcommand{\twoboxes}[2]{\begin{tikzpicture}[baseline={([yshift=-2.5pt]current bounding box.center)},scale=0.025,every node/.style={scale=0.9}, every path/.style={line width=0.2mm}]

\draw    (0,60) -- (10,60) ;
\draw    (0,40) -- (10,40) ;
\draw    (0,20) -- (10,20) ;
\draw    (0,0) -- (10,0) ;
\draw    (160,60) -- (170,60) ;
\draw    (160,40) -- (170,40) ;
\draw    (160,20) -- (170,20) ;
\draw    (160,0) -- (170,0) ;

\draw[rounded corners=1] (10,36) rectangle (160,64);
\draw[rounded corners=1] (10,-4) rectangle (160,24);

\node at (85,51) {\scriptsize $#1$};
\node at (85,11) {\scriptsize $#2$};

\end{tikzpicture}}%

\newcommand{\upboxbrau}[1]{\begin{tikzpicture}[baseline={([yshift=-4.0pt]current bounding box.center)},scale=0.025,every node/.style={scale=0.9}, every path/.style={line width=0.2mm}]

\draw    (0,60) -- (10,60) ;
\draw    (0,40) -- (10,40) ;
\draw    (0,20) -- (70,20) ;
\draw    (0,0) -- (70,0) ;
\draw    (160,60) -- (170,60) ;
\draw    (160,40) -- (170,40) ;
\draw    (100,20) -- (170,20) ;
\draw    (100,0) -- (170,0) ;

\draw[rounded corners=1] (10,36) rectangle (160,64);

\node at (85,51) {\scriptsize $#1$};

\draw    (100.10,-0.07) .. controls (90,3) and (90,17) .. (100.10,20) ;

\draw    (69.90,-0.07) .. controls (80,3) and (80,17) .. (69.90,20) ;

\end{tikzpicture}}%

\newcommand{\downboxbrau}[1]{\begin{tikzpicture}[baseline={([yshift=-1.15pt]current bounding box.center)},scale=0.025,every node/.style={scale=0.9}, every path/.style={line width=0.2mm}]

\draw    (0,60) -- (70,60) ;
\draw    (0,40) -- (70,40) ;
\draw    (0,20) -- (10,20) ;
\draw    (0,0) -- (10,0) ;
\draw    (100,60) -- (170,60) ;
\draw    (100,40) -- (170,40) ;
\draw    (160,20) -- (170,20) ;
\draw    (160,0) -- (170,0) ;

\draw[rounded corners=1] (10,-4) rectangle (160,24);

\node at (85,11) {\scriptsize $#1$};

\draw    (100.10,39.93) .. controls (90,43) and (90,57) .. (100.10,60) ;

\draw    (69.90,39.93) .. controls (80,43) and (80,57) .. (69.90,60) ;

\end{tikzpicture}}%

\newcommand{\lp}{\left(}
\newcommand{\rp}{\right)}

\newcommand{\Up}{\Uparrow}
\newcommand{\Down}{\Downarrow}

\newcommand{\bigotensor}{\mathop{\bigotimes}\limits}

\newcommand{\pone}{\bm{+^{14;23}}}
\newcommand{\ptwo}{\bm{+^{12;34}}}
\newcommand{\pthree}{\bm{+^{1234}}}

\newcommand{\kket}[1]{\left\Vert #1 \right\rangle}
\newcommand{\bbra}[1]{\left\langle #1 \right\Vert}
\newcommand{\bbrakket}[2]{\left\langle #1 \right. \left\Vert #2 \right\rangle}

\begin{document}

\title{Error accumulation in dissipative quantum circuits}

\author{Nadir Samos S{\'a}enz de Buruaga}
\email{nadir.samos@uam.es}
\affiliation{CIAF,Departamento de Física Teórica, Universidad Aut\'onoma de Madrid, C. Francisco Tom\'as y Valiente 7, 28049 Madrid, Spain}
\affiliation{CeFEMA, LaPMET, Instituto Superior Técnico,
Universidade de Lisboa, Av. Rovisco Pais, 1049-001 Lisboa, Portugal.}
\author{Rodrigo M. C. Pereira}
\affiliation{CeFEMA, LaPMET, Instituto Superior Técnico,
Universidade de Lisboa, Av. Rovisco Pais, 1049-001 Lisboa, Portugal.}
\affiliation{Institute of Computer Architecture and Computer Engineering, University of Stuttgart, Germany.}
\affiliation{Center for Integrated Quantum Science and Technology, Germany.}
\author{Karol {\.Z}yczkowski}
\affiliation{Faculty of Physics, Astronomy and Applied Computer Science, Jagiellonian University, ul. \L{}ojasiewicza 11, 30-348 Krak{\'o}w, Poland.}
\affiliation{Center for Theoretical Physics, Polish Academy of Sciences, Al. Lotnik{\'o}w 32/46, 02-668 Warszawa, Poland.}
\author{Sergey Denisov}
\affiliation{Department of Computer Science, OsloMet – Oslo Metropolitan University, 0130 Oslo, Norway.}
\affiliation{NorQSoft - Norwegian Quantum Software Center, Kristian Augusts gate 23, 0164 Oslo, Norway}
\author{Pedro Ribeiro}
\affiliation{CeFEMA, LaPMET, Instituto Superior Técnico,
Universidade de Lisboa, Av. Rovisco Pais, 1049-001 Lisboa, Portugal.}

\maketitle

\begin{abstract}
 Present-day quantum processors are open systems in which dissipation and decoherence degrade the information carried by a quantum state as it propagates through a circuit. We study this degradation in dissipative random quantum circuits, modeling each two-qubit gate as a \emph{diluted unitary} that interpolates between the intended unitary operation and a random dissipative quantum channel. Using the fidelity between the ideal and noisy output states as a diagnostic, we show that sufficiently random unitary gates or dissipative Kraus operators lead to a universal decay of the average fidelity. This decay is determined only by the dissipation strength, system size, and circuit depth, while microscopic details of the gates and noise, including the Kraus rank, appear only in subleading corrections to higher moments. We derive a closed-form expression for the average fidelity in terms of three physically meaningful error parameters, dissipation strength, coherent two-qubit gate-error strength, and connectivity-error probability. Finally, we identify the cases when dissipative effects can be reproduced by an unitary-noise model and when dissipation-induced decoherence remains distinguishable from coherent noise at the level of the average fidelity.
\end{abstract}

\section{Introduction}
By now, quantum computing has moved from the level of abstract models of information processing~\cite{NielsenQuantumComputationQuantum2012} to experimental platforms operating in the noisy intermediate-scale quantum (NISQ) regimes~\cite{preskillQuantumComputingNISQ2018}, where limited connectivity, imperfect gates, finite coherence times, and measurement noise constrain circuit depth while still allowing tests of quantum advantage and quantum simulation~\cite{aaronsonComplexityTheoreticFoundationsQuantum2016,boixoCharacterizingQuantumSupremacy2017,harrowQuantumComputationalSupremacy2017,aruteQuantumSupremacyUsing2019,panSolvingSamplingProblem2022,kretschmerDemonstratingUnconditionalSeparation2025,zhongQuantumComputationalAdvantage2020,zhaoLeapfroggingSycamoreHarnessing2024,lloydUniversalQuantumSimulators1996,georgescuQuantumSimulation2014}. These errors have more than one physical origin, with coherent unitary mismatches, such as over-rotations or miscalibrated two-qubit gates, coexisting with open-system processes produced by coupling to uncontrolled degrees of freedom, including decoherence, dissipation, leakage, and measurement backaction. A useful approach to error accumulation should therefore reveal which mechanisms dominate the degradation of quantum information, when distinct errors become operationally indistinguishable, and which diagnostics can guide benchmarking, mitigation, and, ultimately, error correction.

Random quantum circuits provide a controlled setting in which this approach can be developed and tested. They connect many-body scrambling, random matrix theory, unitary designs, benchmarking protocols, and quantum-chaotic behavior within a common framework~\cite{fisherRandomQuantumCircuits2023,mehtaRandomMatrices2004}. They also underpin quantum-advantage tests~\cite{aaronsonComplexityTheoreticFoundationsQuantum2016,boixoCharacterizingQuantumSupremacy2017}, benchmarking and characterization of quantum processors~\cite{emersonScalableNoiseEstimation2005,knillRandomizedBenchmarkingQuantum2008,magesanScalableRobustRandomized2011,magesanCharacterizingQuantumGates2012,crossScalableRandomisedBenchmarking2016,mollQuantumOptimizationUsing2018,crossValidatingQuantumComputers2019,heinrichRandomizedBenchmarkingRandom2023,mckayBenchmarkingQuantumProcessor2023,morvanPhaseTransitionsRandom2024}, studies of entanglement growth, operator  and magic spreading~\cite{nahumQuantumEntanglementGrowth2017,beraGrowthGenuineMultipartite2020,nahumOperatorSpreadingRandom2018,khemaniOperatorSpreadingEmergence2018,turkeshiMagicSpreadingRandom2025}, unitary $t$-designs~\cite{brandaoLocalRandomQuantum2016,haferkampRandomQuantumCircuits2022}, black-hole information dynamics~\cite{haydenBlackHolesMirrors2007,maganRandomCircuitsBlack2025}, quantum computational complexity~\cite{brandaoModelsQuantumComplexity2021,chapmanQuantumComputationalComplexity2022}, and quantum chaos~\cite{chanSpectralStatisticsSpatially2018,chanSolutionMinimalModel2018,bertiniRandomPermutationCircuits2025}.
They are also central to the practical problem of quantifying how errors accumulate in finite-depth quantum computations. A circuit of $N$ faulty gates with an independent error probability $\epsilon$ has a nominal error-free probability $(1-\epsilon)^N\simeq e^{-\epsilon N}$, but this estimate does not determine which microscopic features of the errors remain visible in many-body observables. Understanding that question is important for benchmarking, error mitigation, and quantum error correction~\cite{caiQuantumErrorMitigation2023,moskeRandomMatrixPerspective2025,terhalQuantumErrorCorrection2015,acharyaQuantumErrorCorrection2025}. Most random-circuit analyses of error accumulation start from unitary imperfections, where the implemented gate differs coherently from the target gate or the circuit connectivity results in qubit permutations different from the intended ones. In Ref.~\cite{samosFidelityDecayError2025}, such errors were analyzed in connectivity-agnostic random quantum circuits using the state fidelity between the ideal and faulty outputs~\cite{guFidelityApproachQuantum2012,goussevLoschmidtEcho2012,samossaenzdeburuagaComparingQuantumComplexity2025,bistronBenchmarkingQuantumDevices2026}.

The dissipative counterpart requires completely positive trace-preserving maps and has become an active interface between open-system dynamics, random matrix theory, and quantum information~\cite{rivasOpenQuantumSystems2012}. Measurements and environment-induced noise have revealed new dynamical phenomena, including measurement-induced phase transitions~\cite{skinnerMeasurementInducedPhaseTransitions2019,potterEntanglementDynamicsHybrid2022} and deep thermalization~\cite{cotlerEmergentQuantumState2023,bejanMatchgateCircuitsDeeply2025}.

In parallel, dissipative quantum chaos has been studied through the spectral and steady-state properties of Lindblad dynamics~\cite{saLindbladianDissipationStronglycorrelated2022,costaSpectralSteadystateProperties2023}. Dissipative and monitored circuits have further revealed connections between non-unitary dynamics, scrambling, and noise-induced structure~\cite{wangNoiseinducedBarrenPlateaus2021,saIntegrableNonunitaryOpen2021,dalzellRandomQuantumCircuits2024,saSpectralTransitionsUniversal2020,liStatisticalMechanicsMonitored2023,yoshimuraRobustnessQuantumChaos2024,pereiraDissipationInducedThresholdIntegrability2025,woldExperimentalDetectionDissipative2025}.


Here We address the following question: To what extent does the microscopic structure of local dissipative noise influence the accumulation of errors in a random circuit? Partial answers to this question have recently been proposed. In Ref.~\cite{dalzellRandomQuantumCircuits2024}, the authors considered output probabilities over bitstrings, using the linear cross-entropy benchmark, and showed that sufficiently deep random circuits can transform local noise into an effective global white-noise model. In contrast, Ref.~\cite{abadUniversalFidelityReduction2022a}studied the effect of a single circuit layer subject to weak Lindblad dissipation, demonstrating that the average fidelity is independent of the specific unitary being implemented. Here, we revisit this question while addressing two important limitations of these previous approaches: the basis dependence of output-probability diagnostics and the restriction to very shallow or perturbative circuit regimes. To address this, we analyze the basis-independent many-body state fidelity and go beyond the perturbative limit, investigating how error accumulation depends on circuit depth and on more general forms of local dissipation.

We consider circuits in which each two-qubit gate is replaced by a \emph{diluted unitary}: a convex combination of the intended unitary channel and a dissipative Kraus map. The dilution parameter $\kappa$ controls the probability weight of the dissipative component, while the Kraus rank $r$ controls the number of dissipative environmental channels. This model is flexible enough to interpolate between unitary random circuits and fully dissipative random maps, while remaining analytically tractable in a random-matrix approximation.

The main result is an explicit formula for the ensemble-averaged fidelity, Eq.~\eqref{eq:universal_fid}. The derivation is exact for a solvable version of the model in which the random qubit permutations are replaced by Haar-random global unitaries. Numerical simulations then show that the same expression accurately describes the original local random circuit after modest depth, regardless of the details of the two-qubit gates and the Kraus operators. Within this regime, the mean fidelity depends on $\kappa$, the number of qubits $L$, and the depth of the circuit $T$, but not on the Kraus rank $r$. We show that the role of the rank is subleading, and that whenever either the two-qubit gates or the Kraus operators are sufficiently random, rank controls only corrections to the concentration of the fidelity around the mean.

We also include unitary gate noise and connectivity errors, obtaining a closed expression, Eq.~\eqref{eq:fid_diluted3param}, that combines three error parameters: dissipative strength $\kappa$, coherent unitary two-qubit noise strength $\alpha$, and a permutation failure probability $p$. This expression identifies a regime in which dissipative and unitary errors are indistinguishable at the level of the average fidelity, as well as a complementary region in which no effective unitary noise can reproduce the dissipative fidelity decay.

The paper is organized as follows. Section~\ref{sec:setting_scene} defines the fidelity diagnostic, the diluted-unitary channel, and the dissipative random circuit. Section~\ref{sec:universal_decay} derives and tests the universal fidelity decay and discusses higher moments. Section~\ref{sec:nonandunitaryfid} adds unitary gate and connectivity errors and analyzes the effective-unitary description of dissipation. Section~\ref{sec:conclusions} summarizes the implications and limitations of the results.

\section{Setting the scene}
\label{sec:setting_scene}
\subsection{Faulty quantum evolution. Fidelity}
Consider $L$ qubits, with Hilbert-space dimension $d=2^L$, initialized in a pure state $\ket{\psi_0}$. The ideal circuit consists of $T$ unitary layers,
\begin{equation}
    \ket{\Psi}=U_TU_{T-1}\cdots U_1\ket{\psi_0}\ .
    \label{eq:ideal_state}
\end{equation}
The implemented evolution is generally not unitary. We denote noisy objects by a tilde and describe the faulty output by a density matrix $\tilde{\rho}$ obtained from a sequence of completely positive trace-preserving maps $\mathcal{E}_\tau$
\begin{equation}
\begin{split}
    \tilde{\rho}&=\mathcal{E}_T(\mathcal{E}_{T-1}(\ldots\mathcal{E}_1(\ket{\psi_0}\bra{\psi_0})))\\&=\sum_{\mu_1,\ldots,\mu_T=0}^{R}\overset{T}{\overleftarrow{\prod_{\tau=1}}}\hat{K}_{\mu_\tau}\ket{\psi_0}\bra{\psi_0}\overset{T}{\overrightarrow{\prod_{\tau=1}}}\hat{K}^\dagger_{\mu_\tau}\  ,
\end{split}
\label{eq:real_state}
\end{equation}
where we write each channel in its Kraus representation, satisfying the usual constraint
\begin{equation}
    \sum_{\mu=0}^{R}\hat{K}^\dagger_\mu \hat{K}_\mu=\mathds{1}\ ,
    \label{eq:Kraus_condition}
\end{equation}
and $0\leq R\leq d^2-1$ is the Kraus rank.
We quantify the deviation between the ideal pure state and the noisy output by the state fidelity
\begin{equation}
     \mathcal{F}=\mel{\Psi}{\tilde{\rho}}{\Psi}\ .
    \label{eq:fid_definition}
\end{equation}

The calculations below use vectorized notation
\begin{equation}
  \kket{\psi\phi} =\ket{\psi}\otimes\ket{\phi^{T}}\ .
\end{equation}

After averaging the initial state over the computational basis, the fidelity can be written as an overlap in four copies of the Hilbert space. We introduce
\begin{equation}
    \begin{split}
    \kket{\pone}&\equiv\sum_{\bm{n}\bm{m}}\kket{\bm{m}\bm{n}\bm{n}\bm{m}}\\ 
    \kket{\ptwo}&\equiv\sum_{\bm{n}\bm{m}}\kket{\bm{m}\bm{m}\bm{n}\bm{n}}\\ 
    \kket{\pthree}&\equiv\sum_{\bm{m}}\kket{\bm{m}\bm{m}\bm{m}\bm{m}}\ , 
    \end{split}
    \label{eq:def_states4copies}
\end{equation}
where $\bm{m}=m_1,\ldots,m_L$ and $m_i=0,1$. The fidelity is then given by (see Appendix~\ref{app:powers_fid})
\begin{equation}
\begin{split}
    \mathcal{F}&=\frac{1}{d}\bbra{\pone}\\
    &\hspace{2cm}\overset{T}{\overleftarrow{\prod_{\tau=1}}}\left(\sum_{\mu_\tau=0}^r\left(\hat{K}_{\mu_\tau}\otimes\hat{K}^*_{\mu_\tau}\right)\otimes U_{\tau}\otimes U_{\tau}^*\right)\\
    &\hspace{4.5cm}\kket{\pthree}\ ,
    \end{split}
    \label{eq:fidelity_posed}
\end{equation}
This expression is the starting point for all ensemble averages. It separates the ideal evolution, appearing in the last two copies, from the noisy Kraus evolution, appearing in the first two copies.

\subsection{Diluted Unitaries}

We model each faulty gate by a diluted unitary channel~\cite{saSpectralTransitionsUniversal2020,woldSpectraNoisyParameterized2025}. It consists of a unitary branch (which can also be noisy) and a dissipative branch,
\begin{equation}
   \hat{K}_0\equiv\sqrt{1-\kappa}\ U_{\tau}\ ,  \hspace{1cm}\hat{K}_{\mu_\tau\neq0}\equiv\sqrt{\kappa}\ K_{\mu_\tau}\ ,
\end{equation}
where the dissipative Kraus operators must also satisfy $\sum_{\mu=1}^{r}K_\mu^\dagger K_\mu=\mathds{1}$ with $1\leq r\leq d^2-1$. In vectorized form,
\begin{equation}
     \kket{\rho_\tau}=\mathcal{E}_\tau\kket{\rho_{\tau-1}}\ ,
     \label{eq:diluted_evolution}
\end{equation}
with
\begin{equation}
     \mathcal{E}_{\tau}=(1-\kappa) U_\tau \otimes U^*_\tau+\kappa\sum_{\mu_{\tau}=1}^rK_{\mu_{\tau}}\otimes K^*_{\mu_{\tau}},
     \label{eq:def_diluted}
\end{equation}
with $\kappa\in[0,1]$. The limit $\kappa=0$ gives a purely unitary gate, whereas $\kappa=1$ gives a purely dissipative random channel. For $\kappa>0$, the full diluted channel also contains the unitary branch, so the full Choi rank is bounded by $r+1$.

Diluted unitaries have been used to interpolate between unitary random-matrix behavior and non-unitary Ginibre-like spectra in studies of dissipative quantum chaos~\cite{saSpectralTransitionsUniversal2020,ginibreStatisticalEnsemblesComplex1965,woldExperimentalDetectionDissipative2025}. They have also been used as probes of integrability signatures in dissipative dynamics~\cite{pereiraDissipationInducedThresholdIntegrability2025}. Here we use the same family of channels for a different purpose: to ask which features of local dissipative noise survive in fidelity decay.
\par
\subsection{Dissipative Random Quantum Circuits}
\begin{figure}
    \centering
    \includegraphics[width=0.95\linewidth]{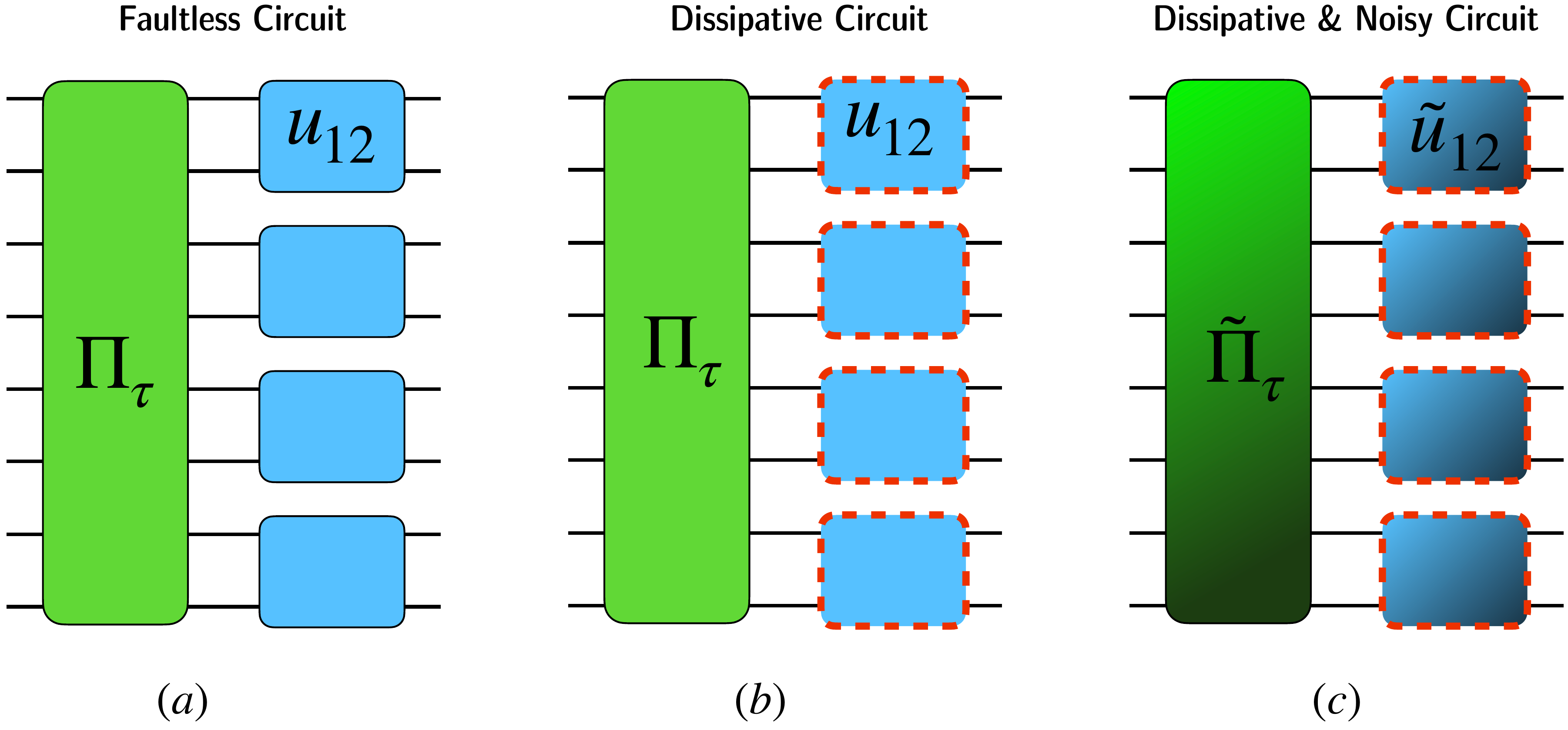}
    \caption{Circuit layer used throughout the paper. (a) The ideal layer is a random permutation $\Pi_\tau$ followed by $L/2$ two-qubit gates $u_{2m-1,2m}$ sampled from CUE(4). (b) In the dissipative model each two-qubit unitary is replaced by a diluted unitary channel with strength $\kappa$. (c) In Sec.~\ref{sec:nonandunitaryfid} we also include unitary two-qubit gate noise and faulty implementations of the permutation.
}
    \label{fig:layer_errors}
\end{figure}
The ideal layer, shown in Fig.~\ref{fig:layer_errors}(a), is
\begin{equation}
    U_\tau= V_\tau \Pi_\tau, \quad V_\tau=\bigotensor_{m=1}^{L/2} u_{2m-1,2m}\ ,
    \label{eq:layer_def}
\end{equation}
where $\Pi_\tau$ implements a random permutation of the $L$ qubits, and the gates $u_{2m-1,2m}$ are independently sampled from the circular unitary ensemble CUE(4). The permutation removes a fixed geometry from the model: after the shuffle, the nearest-neighbor pairings in $V_\tau$ correspond to effectively random pairs of original qubits. Other choices of $\Pi_\tau$ recover more local brickwork or brickwall circuits, while all-to-all permutations model architectures with long-range connectivity~\cite{dahlstenEmergenceTypicalEntanglement2007,oliveiraGenericEntanglementCan2007,emersonConvergenceConditionsRandom2005,harrowRandomQuantumCircuits2009,dalzellRandomQuantumCircuits2022,chapmanQuantumComputationalComplexity2022,sekinoFastScramblers2008}.

We attach dissipation locally to the two-qubit gates, rather than introducing a single Kraus map acting on all $L$ qubits. The resulting noisy layer, represented in Fig.~\ref{fig:layer_errors}(b), is
\begin{equation}
\begin{split}
\mathcal{E}_\tau&=\bigotensor_{m=1}^{L/2}\Bigl[(1-\kappa) u_{2m-1,2m}\otimes u^*_{2m-1,2m}\\
    &+\kappa\sum^r_{\mu_{2m-1,2m}=1} k_{\mu_{2m-1,2m}}\otimes k^*_{\mu_{2m-1,2m}}\Bigr]\\
    &\quad\times(\Pi_\tau\otimes\Pi_\tau)\ ,
    \end{split}
    \label{eq:small_diluted}
\end{equation}
where in this case $1\leq r\leq 15$. This choice keeps the environmental coupling local at the gate level and makes it possible to ask how errors accumulate across a circuit layer by layer.

\section{Universal Fidelity decay}
\label{sec:universal_decay}
In this section, we obtain a universal expression for the fidelity decay of the dissipative RQC presented above. We present the main points of the derivation, relegating the most technical details of the computation to Appendix~\ref{app:universalfid} for the interested reader. 
The fidelity Eq.~\eqref{eq:fidelity_posed} in terms of the map Eq.~\eqref{eq:small_diluted} takes the form
\begin{equation}
\begin{split}
    \mathcal{F}&=\frac{1}{d}\bbra{\pone}
    \overset{T}{\overleftarrow{\prod_{\tau=1}}}\Bigl[ 
   \bigotensor_{m=1}^{L/2}\Bigl((1-\kappa)\bm{u}_{2m-1,2m} \\
   &\ \ +\kappa\sum^r_{\mu_{2m-1,2m}=1} \bm{k}_{\mu_{2m-1,2m}}\Bigr)\bm{\Pi}_\tau\Bigr]\kket{\pthree}\ ,
    \end{split}
\label{eq:fidelity_localdiluted}
\end{equation}
where $\bm{u}_{m,m'}\equiv (u_{m,m'}\otimes u^*_{m,m'})^{\otimes2}$, $\bm{\Pi}_\tau\equiv\Pi_\tau^{\otimes4}$, and
\begin{equation}
    \bm{k}_{\mu_{mm'}}\equiv k_{\mu_{m,m'}}\otimes k^*_{\mu_{m,m'}}\otimes u_{m,m'}\otimes u^*_{m,m'}\ .
    \label{eq:4copies_kraus}
\end{equation}
The exact average of Eq.~\eqref{eq:fidelity_localdiluted} is difficult because the permutations correlate the local gates across successive layers. We therefore first solve a random-matrix version of the problem in which each permutation is replaced by an independent Haar-random unitary, $\Pi_\tau\to R_\tau\in\text{CUE}(d)$. This replacement is not an identity of circuit ensembles: it changes detailed scrambling and entanglement properties. The comparison with the original permutation circuit is then made numerically and, as shown below, exhibits remarkably good agreement even for very shallow circuits, where the underlying approximation is at its crudest.

The Haar average that replaces $\overline{\bm{\Pi}}$ is
$\overline{\bm{\mathcal{R}}}=\overline{R\otimes R^*\otimes R\otimes R^*}$.
We define this replacement map through the orthonormal basis
\begin{equation}
\begin{split}
    \kket{\Up} &=\frac{1}{d} \kket{\ptwo}\\ \kket{\Down}&=\frac{1}{\sqrt{d^2-1}}\left(\kket{\pone}-\frac{1}{d}\kket{\ptwo}\right)\ ,
\end{split}
\label{eq:spinbasis}
\end{equation}
where the four-copy states are given in Eq.~\eqref{eq:def_states4copies}. In this basis,
\begin{equation}
\overline{\bm{\Pi}}\to\overline{\bm{\mathcal{R}}}\equiv\kket{\Up}\bbra{\Up}+\kket{\Down}\bbra{\Down}\ ,
    \label{eq:avg_haar_diag}
\end{equation}
which follows from the standard Weingarten calculation~\cite{collinsWeingartenCalculus2022}.
\begin{figure}[H]
\includegraphics[width=0.90\linewidth]{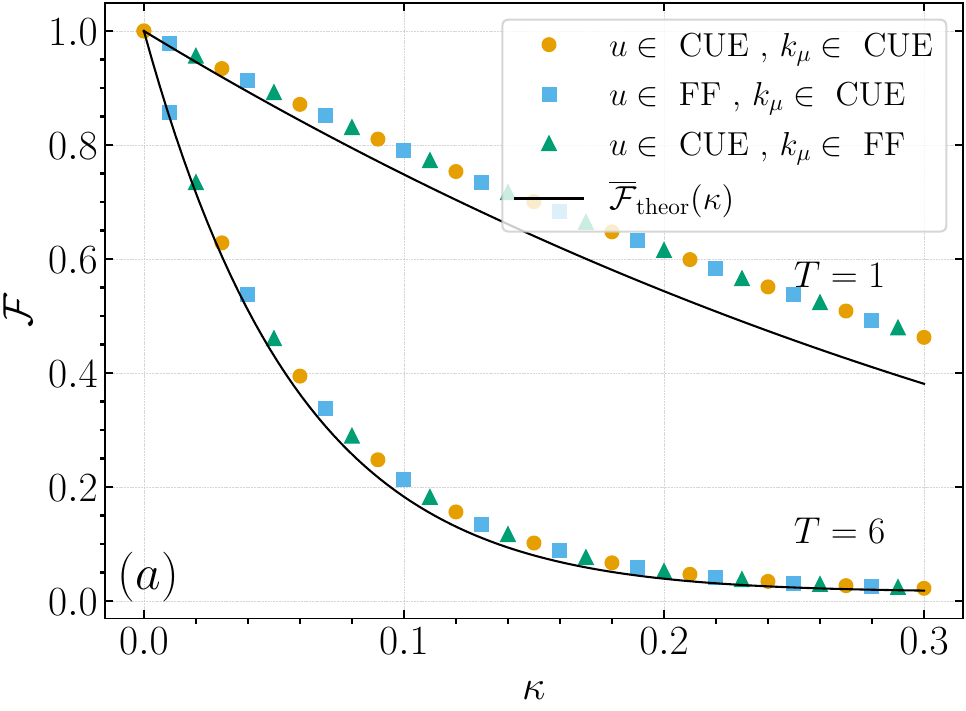}\\
\includegraphics[width=0.90\linewidth]{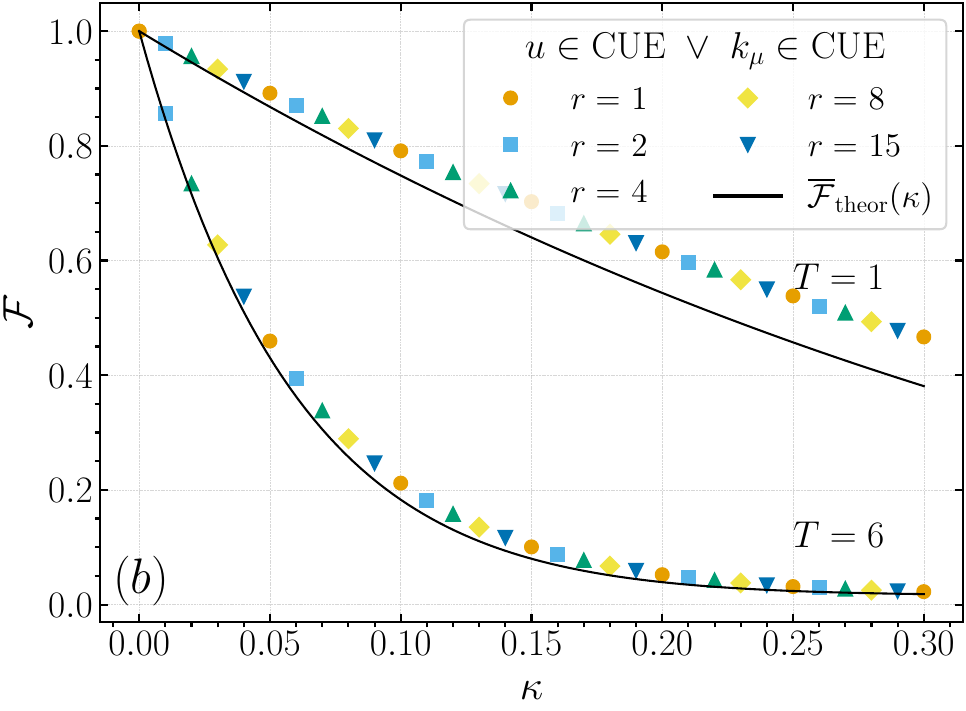}\\
\includegraphics[width=0.90\linewidth]{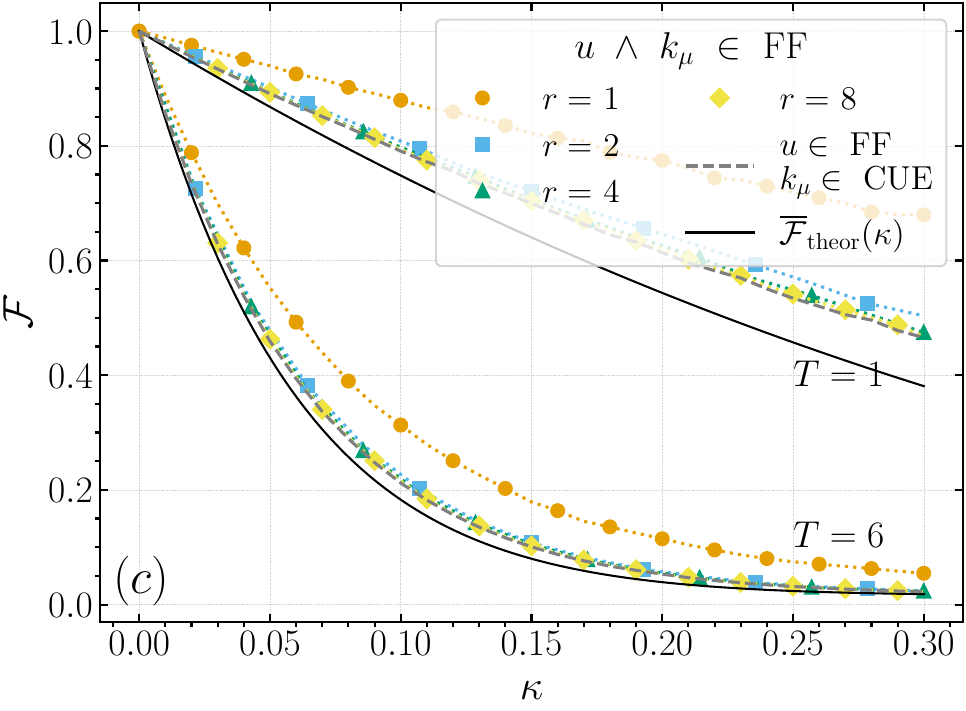}
  \caption{Mean fidelity decay for $L=6$. (a) When either the two-qubit unitaries or the dissipative Kraus operators are sampled from a generic ensemble, the numerical data collapse onto Eq.~\eqref{eq:universal_fid}. (b) In the same scenario than above, the mean fidelity is insensitive to the dissipative Kraus rank $r$. (c) When both the unitaries and the dissipative component are restricted to free-fermion dynamics, a finite-depth rank dependence remains, but the data approach the generic behavior at larger depth given by the gray dashed line. Notice that the random permutations are implemented as fermionic-mode shuffles via fermionic SWAP decompositions.}
\label{fig:universal_fid}
\end{figure}
The remaining local averages are over the two-qubit gates and the dissipative Kraus operators. If either the two-qubit gates are Haar random, $u_{m,m'}\in\text{CUE}(4)$, or the Kraus operators are generated by slicing a block column of a Haar-random unitary of dimension $4r$, the same two-copy contraction is obtained. The detailed calculation carried out in Appendix~\ref{app:universalfid} then gives
\begin{equation}
    \overline{\mathcal{F}}(\kappa)= \frac{1}{d}+\Biggl(\frac{\left(16-15\kappa\right)^{L/2}-1}{d^2-1}\Biggr)^T\Biggl(1-\frac{1}{d}\Biggr)\ .
    \label{eq:universal_fid}
\end{equation}

Equation~\eqref{eq:universal_fid} is the central result for the solvable model. It depends on the dissipative weight $\kappa$, the system size, and the depth, but not on the Kraus rank $r$ of the dissipative component. Thus, although rank-sensitive observables are known to matter in dissipative quantum chaos and integrability diagnostics~\cite{saSpectralTransitionsUniversal2020,pereiraDissipationInducedThresholdIntegrability2025,woldExperimentalDetectionDissipative2025}, the ensemble-averaged state fidelity is blind to $r$ at this level.

Figures~\ref{fig:universal_fid}(a,b) provide direct numerical evidence for these claims. To make this test nontrivial, we must also compare against a non-generic ensemble. We therefore consider free-fermion (matchgate) dynamics, which is integrable and efficiently classically simulable~\cite{valiantQuantumCircuitsThat2012,terhalClassicalSimulationNoninteractingfermion2002,jozsaMatchgatesClassicalSimulation2008}. Figure~\ref{fig:universal_fid}(c), while not a strict test of rank independence itself, shows that even in this structured non-generic case the mean fidelity remains consistent with the same behavior described by Eq.~\eqref{eq:universal_fid} for sufficiently deep circuits.

Overall, the local diluted-unitary construction Eq.~\eqref{eq:small_diluted} combined with random permutations provides sufficient effective randomness that the structured character of the dynamics and dissipation is largely washed out at the level of mean fidelity. Eq.~\eqref{eq:universal_fid} therefore provides an effective universal description of many-body error accumulation.

\subsection{Higher moments of fidelity}
The absence of $r$ from Eq.~\eqref{eq:universal_fid} raises a natural question: does the Kraus rank disappear from the full distribution of fidelities, or only from the mean? Higher moments show that rank dependence is present, but parametrically suppressed.

We start by writing the higher moments of the fidelity $\overline{\mathcal{F}^k}$ as an overlap of states belonging to a $4k$-copy Hilbert space (see App. \ref{app:powers_fid}):

\begin{equation}
\begin{split}
        \mathcal{F}^k=\frac{1}{d}\bbra{\bm{\frown}_{4k}}\overset{T}{\overleftarrow{\prod_{\tau=1}}}\Biggl[\bigotimes_{i=1}^k\Biggl\{ \left(\sum_{\mu^i_\tau=0}^r\hat{K}_{\mu^i_\tau}\otimes\hat{K}^*_{\mu^i_\tau}\right)\\\otimes U_{\tau}\otimes U_{\tau}^*\Biggr\}\Biggr]\kket{\boldsymbol{\perp}_{4k}},
        \end{split}
    \label{eq:powerF}
\end{equation}
where $\kket{\bm{\frown}_{4k}}\equiv\kket{\boldsymbol{+^{1,\ldots,2k\ ;\ 2k+1,\ldots,4k}}}$ and $\kket{\boldsymbol{\perp}_{4k}}\equiv\kket{\bm{+^{1,\ldots,4k}}}$ are the natural extensions of Eq.~\eqref{eq:def_states4copies}.

The exact treatment of Eq.~\eqref{eq:powerF} is considerably more involved than the mean fidelity because the number of Weingarten contractions grows rapidly with $k$. We therefore use an analytically simpler all-to-all diluted-unitary circuit as a guide. Its mean fidelity is
\begin{figure}[t]
    \centering
    \includegraphics[width=0.99\linewidth]{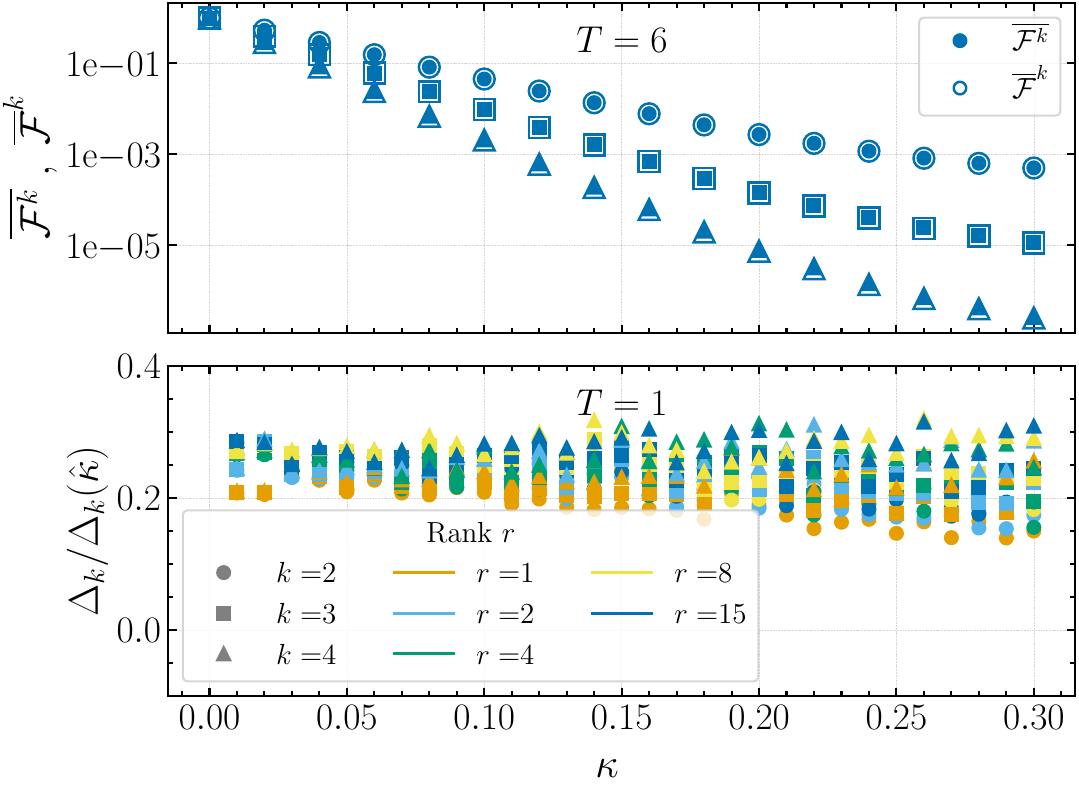}
    \caption{Concentration around the average fidelity. Top: comparison between $\overline{\mathcal{F}^k}$ and $\overline{\mathcal{F}}^k$ as functions of $\kappa$ for $L=6$. Bottom: numerical test of the ansatz Eq.~\eqref{eq:ansatz_kvariance}; for one layer, $g(r,1)=1/r$ collapses the data for the ranks and moments shown. For generic depth, $g(r,T)$ is unknown and captures the imperfections of the ansatz.}
    \label{fig:kvariance1layer}
\end{figure}
\begin{equation}
\overline{\mathcal{F}}_{\mathrm{AtoA}}=\frac{1}{d}+(1-\kappa)^T\left(1-\frac{1}{d}\right)\ ,
    \label{eq:avgfid_alltoall}
\end{equation}
which is the same expression obtained in Ref.~\cite{escofetAccurateEfficientAnalytic2025a} for the application of $T$ depolarizing channels. Its functional form, together with Eq.~\eqref{eq:universal_fid}, motivates the effective dissipation parameter
\begin{equation}
\begin{split}
       \hat{\kappa}(\kappa,L)&\equiv \frac{d^2-(16-15\kappa)^{L/2}}{d^2-1}\\
       &\approx 1-\left(1-\frac{15}{16}\kappa\right)^{L/2\ }. 
\end{split}
    \label{eq:kappahat}
\end{equation}

\begin{figure*}[t]
    \centering
    \includegraphics[width=1\textwidth]{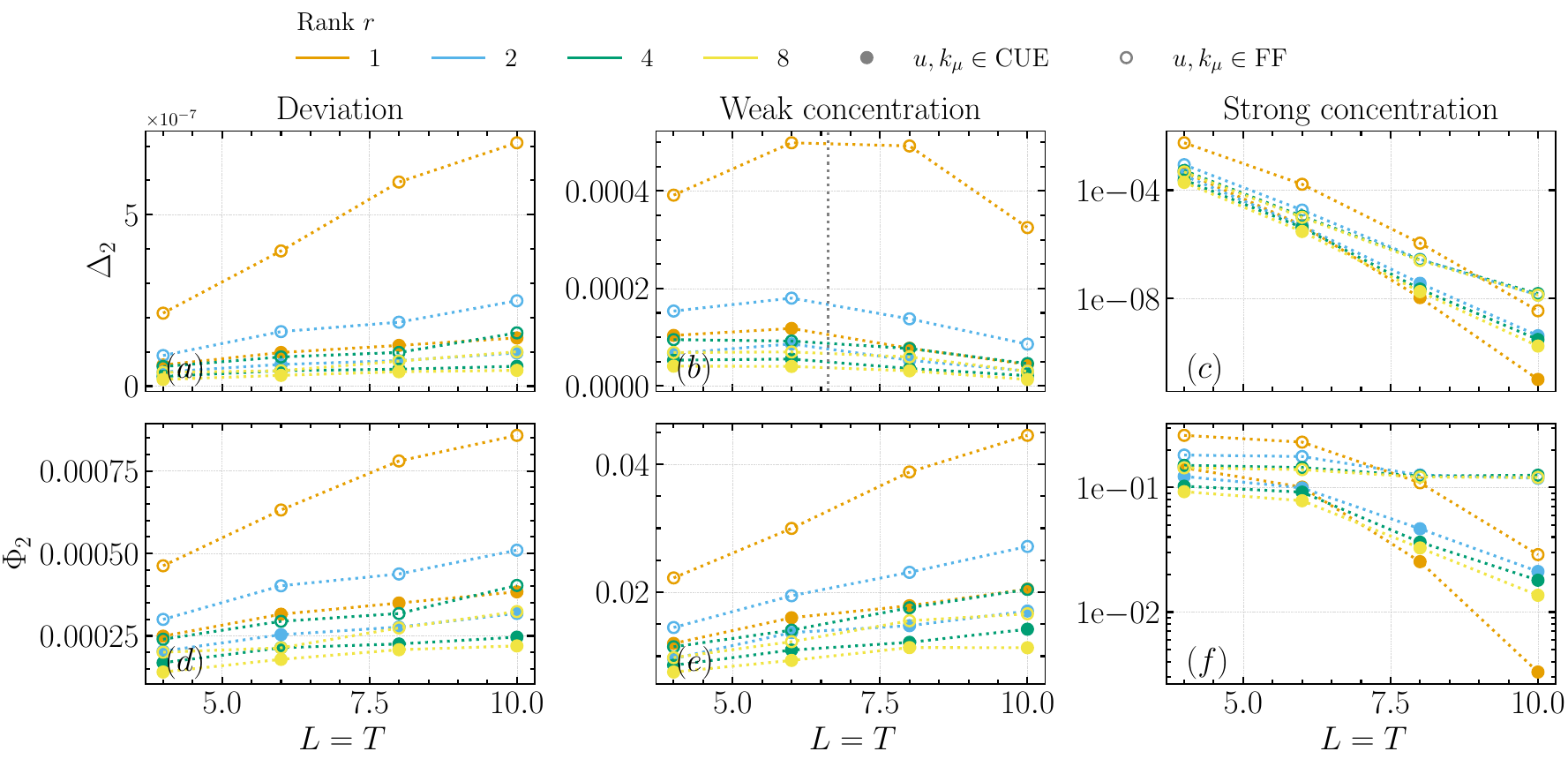}
\caption{Scaling with system size of $\Delta_2$ and $\Phi_2$ for square circuits, $T=L$. The three columns show weak, intermediate, and strong dissipation ($\kappa=0.00005,\ 0.025, \ 0.3$ respectively). In the weak case, $\Delta_2$ and $\Phi_2$ grow over the accessible sizes. In the intermediate case, $\Delta_2$ is already small while the relative fluctuation $\Phi_2$ still grows. In the strong case, both quantities decrease with system size. The dashed line in panel (b) marks the prediction for $L_{\mathrm{max}}$ from Eq.~\eqref{eq:Lmax}.}
\label{fig:concentration}
\end{figure*}

 We use the all-to-all moment calculation, summarized in Appendix~\ref{app:highermoments}, to motivate the scaling form for the local RQC.

Figure~\ref{fig:kvariance1layer}(a) shows that $\overline{\mathcal{F}^k}$ closely follows $\overline{\mathcal{F}}^k$, indicating concentration. We quantify the absolute deviation by
\begin{equation}
\Delta_k\equiv\overline{\mathcal{F}^k}-\overline{\mathcal{F}}^k\ ,
\label{eq:def_kvariance}
\end{equation}
and the relative fluctuation by
\begin{equation}
    \Phi_k\equiv\frac{(\Delta_k)^{1/k}}{\overline{\mathcal{F}}}\ ,
    \label{eq:def_kvariation}
\end{equation}
Strong concentration requires both quantities to vanish with increasing system size.

As we show in Appendix~\ref{app:highermoments}, the leading term of the k-variance for the all-to-all circuit is
\begin{equation}
    \Delta_k^{\textrm{AtoA}}\sim \frac{T}{d^2r}\binom{k}{2}(1-\kappa)^{k T-2}\kappa^2\ .
    \label{eq:kvariance_atoa}
\end{equation}
The rank dependence is explicit, but it is multiplied by the concentration factor $1/d^2$. For the local RQC we use the same functional form with $\kappa$ replaced by $\hat{\kappa}$ and collect the remaining rank and depth dependence in a prefactor:
\begin{equation}
\Delta_k\sim \frac{g(r,T)}{16}\binom{k}{2}(1-\hat{\kappa})^{k T-2}\hat{\kappa}^2,
\label{eq:ansatz_kvariance}
\end{equation}
For one layer, consistency with the all-to-all result gives $g(r,1)=1/r$. The collapse in Fig.~\ref{fig:kvariance1layer} supports this scaling for the ranks and moments shown.

\subsubsection{Absolute concentration}
Unlike Eq.~\eqref{eq:kvariance_atoa}, the local-circuit ansatz predicts a nonmonotonic dependence on $L$ at fixed depth. The maximum occurs near
\begin{equation}
    L_\text{max}=2\frac{\log\left(1-\frac{2}{k T}\right)}{\log\left(1-\frac{15}{16}\kappa\right)}\sim \frac{64}{15k T\kappa}\ .
    \label{eq:Lmax}
\end{equation}
Equivalently, the dissipation scale at which the preasymptotic growth ends is
\begin{equation}
    \kappa_\Delta\sim1-\left(1-\frac{2}{kT}\right)^{2/L}\ . 
\end{equation}
For $\kappa<\kappa_\Delta$, the accessible sizes lie in a weak-dissipation regime where $\Delta_k$ initially grows as $\Delta_k\propto (15\kappa L/32)^2$. This behavior is visible in Fig.~\ref{fig:concentration}(a). At intermediate dissipation, the maximum appears within the simulated range, Fig.~\ref{fig:concentration}(b), while for stronger dissipation, the absolute deviation decreases rapidly with system size, Fig.~\ref{fig:concentration}(c); for square circuits ($T=L$) this decay scales as $\Delta_k\sim(1-\kappa)^{kL^2/2}$, faster than the all-to-all estimate $\Delta_k^{\textrm{AtoA}}\sim(1-\kappa)^L4^{-L}$.

\subsubsection{Relative concentration}
The relative fluctuation $\Phi_k$ can behave differently from the absolute deviation. In the regime $2^{-L}\ll(1-\hat{\kappa})^T$, the ansatz gives
\begin{align}
 \Phi_k&\sim f^{1/k}(r,T)\left(\left(1-\frac{15}{16}\kappa\right)^{-L/2}-1\right)^{2/k}\nonumber\\&\sim f^{1/k}(r,T)\exp\left(\frac{15L\kappa}{16 k}\right)\ ,\\
 &\text{with}\  f(r,T)=\Biggl(\frac{g(r,T)}{16}\binom{k}{2}\Biggr)^{1/k}\nonumber.
\end{align}
The crossover occurs when $2^{-L}=(1-\hat{\kappa})^T$, giving
\begin{equation}
    \kappa_\phi=\frac{15}{16} \Bigl(1-\frac{1}{4^{1/T}}\Bigr),
\end{equation}
 
Typically $0<\kappa_\Delta<\kappa_\phi<1$. Thus, there is an intermediate regime in which the fidelity is already absolutely concentrated, while relative fluctuations still grow with $L$; we call this weak concentration. For larger $\kappa$, both $\Delta_k$ and $\Phi_k$ decrease with system size, giving strong concentration. The empty markers in Fig.~\ref{fig:concentration} show that the fully free-fermion ensemble exhibits the same qualitative regimes, although finite-size rank effects are more visible than in the generic ensemble.

\section{Error accumulation with unitary errors and dissipation}
\label{sec:nonandunitaryfid}
We now add the two unitary error mechanisms studied in Ref.~\cite{samosFidelityDecayError2025}: faulty implementations of the permutation layer and unitary two-qubit gate errors. The model is shown schematically in Fig.~\ref{fig:layer_errors}(c). The goal is to determine how these errors combine with the dissipative strength $\kappa$ in the mean fidelity.

Following Ref.~\cite{samosFidelityDecayError2025}, the intended permutation $P$ is decomposed into SWAP operations. Each required SWAP fails with probability $p\in[0,1]$, producing a faulty permutation $\tilde{P}(p)$ and the corresponding operator $\tilde{\Pi}(p)$. Coherent two-qubit noise is modeled by
\begin{equation}
    \tilde{u}_{m,m'} = e^{i\alpha h_{m,m'}} u_{m,m'}\ ,
    \label{eq:unitary_noise}
\end{equation} 
where $h_{m,m'}$ is drawn from the Gaussian Unitary Ensemble (GUE), and $\alpha \geq 0$ controls the noise strength. 

For $\kappa=0$, the corresponding average fidelity was obtained in Ref.~\cite{samosFidelityDecayError2025} by embedding each faulty permutation between two Haar-random unitaries, $\Pi_\tau\to R_{1,\tau}\Pi_\tau R_{2,\tau}$. In our notation that result reads
\begin{equation}
\begin{split}
    \overline{\mathcal{F}} & =  \frac{1}{d} \Bigl( 1 + \Bigl[\lp\frac{\delta(p)-1}{d^2-1}\rp \\
    & \times \frac{\lp4+12f(\alpha)\rp^{L/2}-1}{d^2-1} \Bigr]^T (d-1) \Bigr)\ ,
    \end{split}
    \label{eq:f_unitarycase}
\end{equation}
where $\delta(p)=\overline{\tr^2\tilde{\Pi}^T\Pi}$ is the second moment of the number of invariant computational-basis states under the relative permutation $\tilde{P}(p)P^{-1}$. For small $p$ in a fully connected architecture, Ref.~\cite{samosFidelityDecayError2025} gives
$$\delta(p)\approx 4^Le^{-p(L+\gamma-\log L)}\ ,$$
where $\gamma$ is Euler's constant. Thus $p=0$ corresponds to a correctly implemented permutation.
\par
The function $f(\alpha)$ is the Fourier transform of the two-point function of the GUE and, for two-qubit gates, is~\cite{delcampoScramblingSpectralForm2017,samosFidelityDecayError2025}
\begin{equation}
    \begin{split}
    f(\alpha)= e^{-\alpha ^2} \Bigl(&1-4\alpha^2+\frac{23}{6}\alpha^4-2\alpha^6\\
    &+\frac{25}{72}\alpha^8-\frac{1}{36}\alpha ^{10}\Bigr)\ ,
    \end{split}
    \label{eq:falpha}
\end{equation}
For small $\alpha$ one may use $f(\alpha)\simeq e^{-5\alpha^2}$.
 
Appendix~\ref{app:fidunitdis} extends the same calculation to the diluted-unitary channel. The averaged faulty-permutation block is
\begin{equation}
\overline{\bm{\mathcal{R}}\bm{\Pi}\bm{\mathcal{R}}}=\kket{\Up}\bbra{\Up}+\frac{\delta(p)-1}{d^2-1}\kket{\Down}\bbra{\Down},
\end{equation}

\par

while the local unitary gate noise gives
\begin{align}
    \overline{\bm{u}_{2m-1,2m}} &=
     \kket{\Up_{2m-1,2m}}\bbra{\Up_{2m-1,2m}} \nonumber \\
     &+\frac{4f(\alpha)+1}{5}\kket{\Down_{2m-1,2m}}\bbra{\Down_{2m-1,2m}}\ ,
\end{align}

and the averaged Kraus contribution is
\begin{equation}
    \overline{\bm{k}_{m,m'}}= \overline{k_{\mu_{m,m'}}\otimes k^*_{\mu_{m,m'}}}\otimes \overline{u_{m,m'}\otimes u^*_{m,m'}}\ .
\end{equation}
The unitary factor has only one copy,
\begin{equation}
    \overline{u_{m,m'}\otimes u^*_{m,m'}}=\frac{1}{4}\kket{\bm{+^{34}_{mm'}}}\bbra{\bm{+^{34}_{mm'}}}\ ,
\end{equation}
whereas a Kraus operator generated from a Haar-random Stinespring unitary contributes
\begin{equation}
    \overline{k_{\mu_{m,m'}}\otimes k^*_{\mu_{m,m'}}}=\frac{1}{4r}\kket{\bm{+^{12}_{mm'}}}\bbra{\bm{+^{12}_{mm'}}}, \quad \forall \ \mu_{mm'}.
\end{equation}
Therefore
\begin{equation}
\begin{split}
    \overline{\bm{k}_{m,m'}}&=\frac{1}{16r}\kket{\ptwo_{mm'}}\bbra{\ptwo_{mm'}}\\&=\frac{1}{r}\kket{\Up_{mm'}}\bbra{\Up_{mm'}}\ ,
    \end{split}
    \label{eq:avg_ofkraus}
\end{equation}
where the last equality uses Eq.~\eqref{eq:spinbasis}. 

Assembling the averaged blocks gives

\begin{equation}
\begin{split}
    \overline{\mathcal{F}} & =  \frac{1}{d}\Bigl(1+\Bigl[\lp\frac{\delta(p)-1}{d^2-1}\rp \\
    & \times \frac{\lp4+12f(\alpha)(1-\kappa)-3\kappa\rp^{L/2}-1}{d^2-1}\Bigr]^T(d-1)\Bigr)\ .
    \end{split}
\label{eq:fid_diluted3param}
\end{equation}

\begin{figure}
    \centering
    \includegraphics[width=0.85\linewidth]{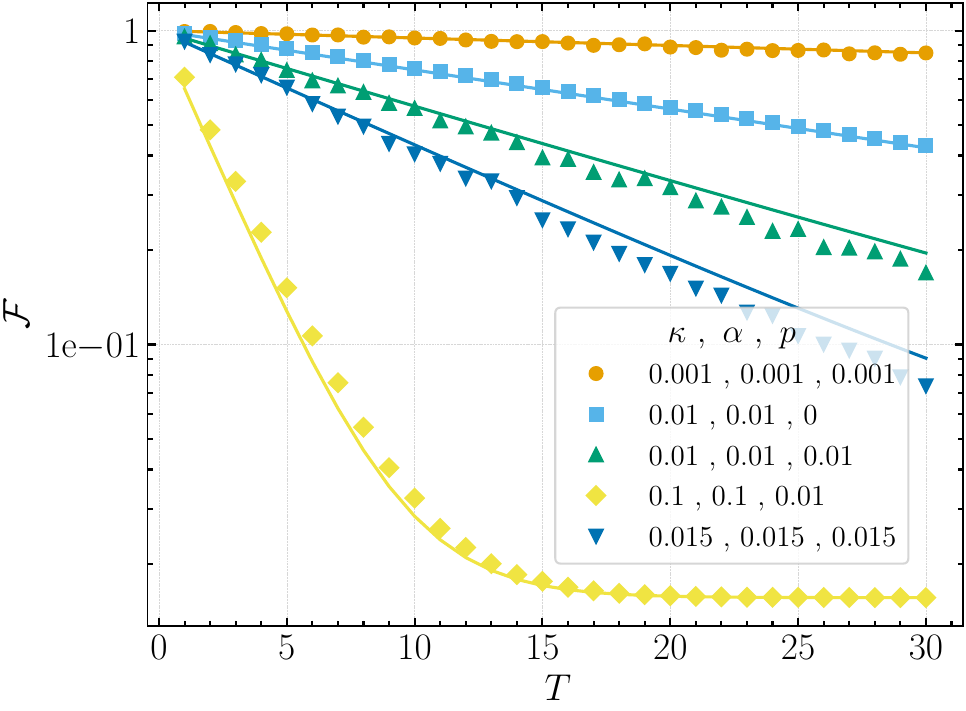}
    \caption{Average fidelity of circuit of $L=6$ as a function of depth $T$ for models with rank $r=1$ for several values of unitary noise $\alpha$, faulty permutations $p$ and dissipation strength $\kappa$. Solid lines are given by Eq.~\eqref{eq:fid_diluted3param}.}
    \label{fig:3errors}
\end{figure}

For $\kappa=0$ this reduces to Eq.~\eqref{eq:f_unitarycase}.  For $\kappa=1$ the average fidelity is $1/d$, independently of $\alpha$ and $p$, because the output has lost all memory of the ideal unitary branch. Figure~\ref{fig:3errors} compares Eq.~\eqref{eq:fid_diluted3param} with numerical data for several combinations of the three error parameters.
\par
\subsection{Effective unitary noise}
Finally, set $p=0$ and ask whether the effect of dissipation on the average fidelity can be reproduced by a purely unitary circuit with an effective noise strength $\alpha_{\rm eff}$. Equating Eq.~\eqref{eq:f_unitarycase} to Eq.~\eqref{eq:fid_diluted3param} gives
\begin{equation}
    f(\alpha)(1-\kappa)-\frac{\kappa}{4}-f(\alpha_{\text{eff}})=0\ ,
     \label{eq:condition_alpha}
\end{equation}
which is independent of $L$ and $T$. In the perturbative regime, $f(\alpha)\simeq1-5\alpha^2$, and
\begin{figure}[H]
    \centering
    \includegraphics[width=0.90\linewidth]{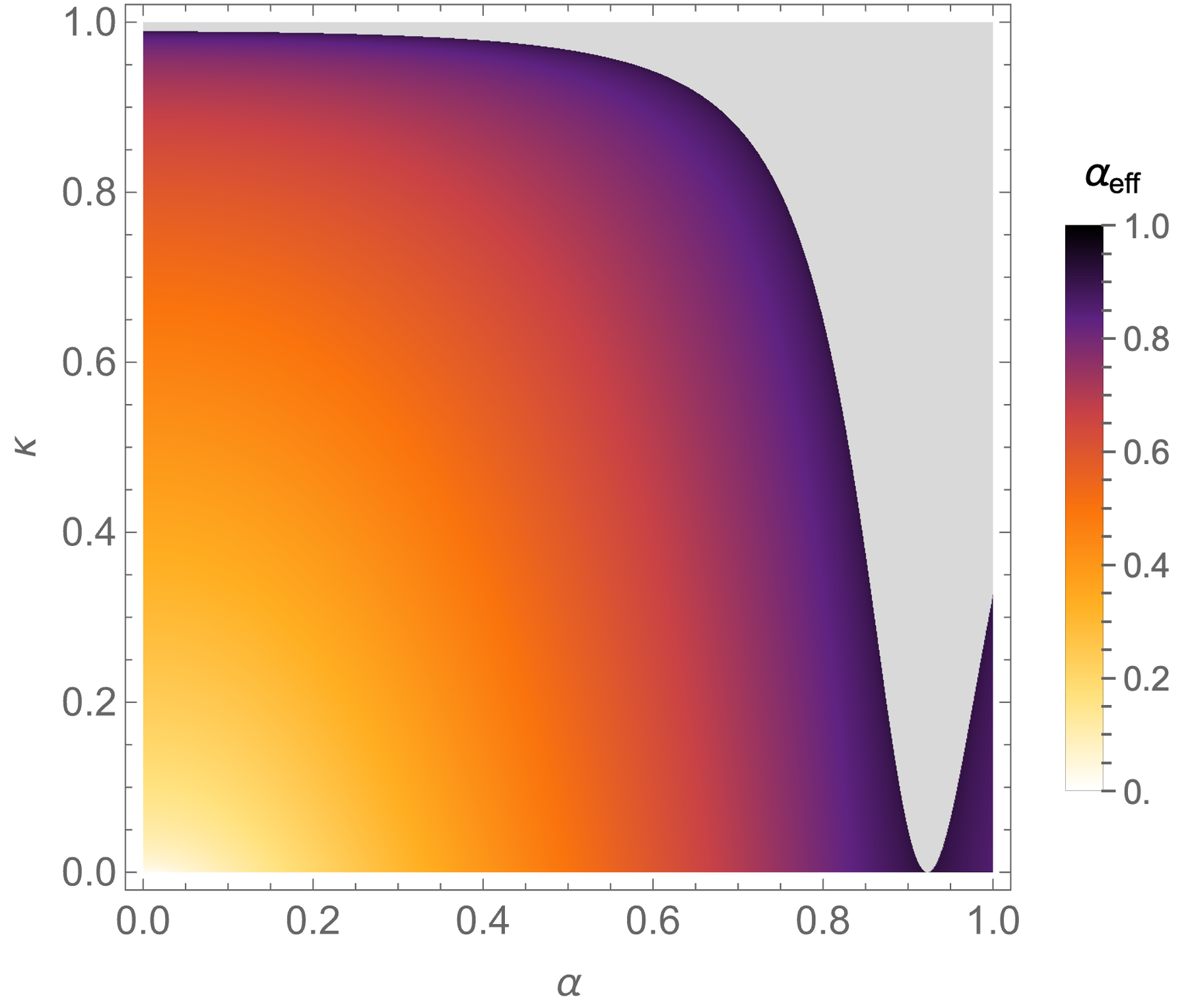}
        \includegraphics[width=0.90\linewidth]{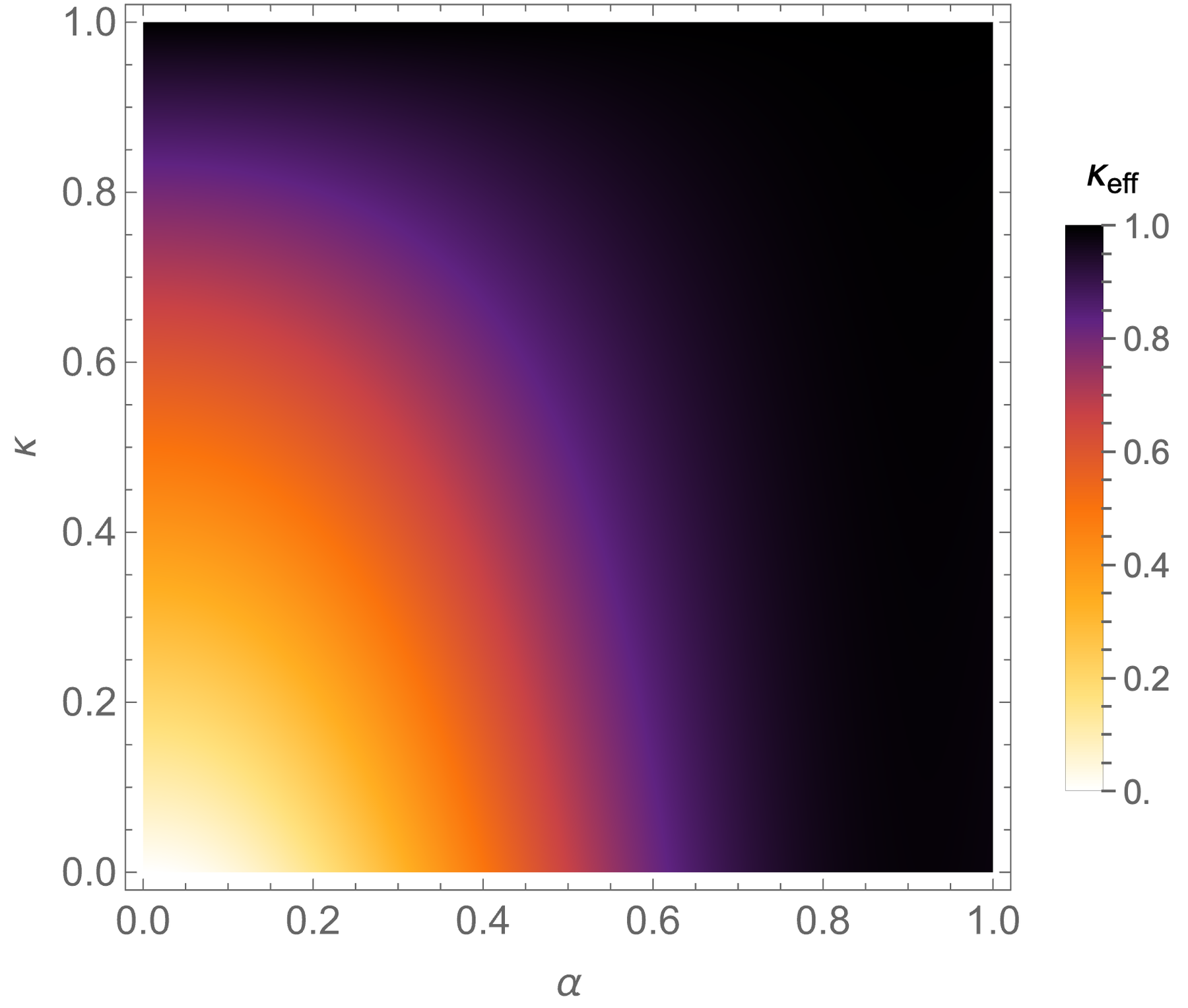}
    \caption{Effective-noise maps obtained from the full condition Eq.~\eqref{eq:condition_alpha}. Top: inferred unitary noise $\alpha_{\rm eff}$ for a circuit with unitary noise $\alpha$ and dissipation $\kappa$; the gray region has no unitary-noise solution. Bottom: inferred effective dissipation for a purely unitary noisy circuit.}
    \label{fig:range_validity}
\end{figure}
\begin{equation}
    \alpha^2_{\rm eff}=\alpha^2+\frac{\kappa}{4}-\alpha^2\kappa \, .
    \label{eq:alphaeff}
\end{equation}

Equation~\eqref{eq:alphaeff} shows that, at the level of the average fidelity and for weak errors, unitary noise and dissipation are not separately identifiable. Additional observables are therefore required to distinguish these mechanisms.

\par
The perturbative formula is reliable only while both $\alpha$ and $\alpha_{\rm eff}$ are small. Solving Eq.~\eqref{eq:condition_alpha} with the full expression for $f(\alpha)$ gives the broader picture in Fig.~\ref{fig:range_validity} (Top). A purely unitary noisy circuit can reproduce the dissipative average fidelity over a wide region of parameter space, but a gray region remains where no real $\alpha_{\rm eff}$ exists. In that regime, dissipation cannot be reinterpreted as an effective unitary error, even for this coarse fidelity diagnostic. On the contrary, finding an effective dissipation parameter $\kappa_{\rm eff}$ that also reproduces the unitary noise is always possible, as it can be seen in Fig.~\ref{fig:range_validity}~(Bottom).

\section{Conclusions}
\label{sec:conclusions}
We have studied fidelity decay in random quantum circuits whose two-qubit gates are replaced by local diluted-unitary channels. In a solvable Haar-substituted model, the ensemble-averaged fidelity is given by Eq.~\eqref{eq:universal_fid}.  The formula depends on the dissipative strength $\kappa$, system size $L$, and depth $T$, but not on the Kraus rank $r$ of the dissipative component. Numerical simulations of the original local circuit show that the same expression describes the mean fidelity whenever either the unitary gates or the Kraus operators are generic, and that structured free-fermion circuits flow toward the same behavior with increasing depth.

The rank independence is specific to the mean fidelity. Higher moments reveal residual $r$ dependence, but the leading corrections are suppressed by concentration factors that decrease with Hilbert-space dimension. Thus the mean fidelity fixes the robust decay scale, whereas fluctuations provide a more sensitive probe of channel-specific structure.

We also derived Eq.~\eqref{eq:fid_diluted3param}, which combines dissipative noise, unitary two-qubit gate noise, and faulty permutations. This expression shows when dissipation can be absorbed into an effective unitary noise parameter and identifies a complementary regime where no such effective description exists. Thus, fidelity-based benchmarks can obscure the physical distinction between unitary and dissipative mechanisms in weak-noise regimes, but they also expose parameter regions where that ambiguity breaks down.

We have extended the unitary-independence result under weak Lindbladian dissipation reported in Ref.~\cite{abadUniversalFidelityReduction2022a} to finite-depth random circuits with local dissipative channels. In the regimes identified here, the mean many-body fidelity is insensitive to microscopic gate and channel details, while nonuniversal information, such as Kraus rank, re-enters only at higher moments. We further provide a complementary universality statement to Ref.~\cite{dalzellRandomQuantumCircuits2024}: while that work establishes white-noise universality at the level of computational-basis bitstring distributions, here we establish universality for basis-independent many-body state fidelity and its moments. These results indicate that effective randomness generated by local diluted-unitary mixing suppresses microscopic structure at the mean-fidelity level, while retaining model-specific signatures in higher-order fluctuations.

The main limitation of the analysis is that the closed-form expressions rely on a random-matrix replacement of the permutation layer and on random-channel assumptions for at least one branch of the diluted gate. The numerical evidence indicates that the resulting formulas are robust for the local circuits studied here, but more structured environments may require additional diagnostics. Natural next steps include studying channel families in which the Kraus rank affects the mean fidelity, comparing fidelity with basis-dependent output statistics~\cite{dalzellRandomQuantumCircuits2024}, and extending the analysis to experimentally reconstructed noise models.

\begin{acknowledgments}
    \par
We are thankful to Rafa{\l} Bistro{\'n} and Marcin Rudzi\-{\'n}ski  for a fruitful collaboration on an earlier project, which triggered this work. 
This work was supported by the Research Council of Norway, project “IKTPLUSS-IKT og digital innovasjon - 333979” (SD), and by FCT-Portugal, Grant Agreement No. 101017733 (PR, NS, and RP), as part of the QuantERA II project “DQUANT: A Dissipative Quantum Chaos perspective on Near-Term Quantum Computing”, DOI. 10.54499/QuantERA/0003/2021. 
PR, NS, and RP acknowledge support by CeFEMA, under contracts UID/04540/2025,UID/PRR/04540/2025, and UID/PRR2/04540/2025 with AI2, DOI: 10.54499/UID/PRR/04540/2025; 10.54499/UID/PRR/04540/2025; and 10.54499/UID/PRR2/04540/2025. 
NS further acknowledges support by AEI through project PID2024-159024NB-C22.
PR further acknowledges support by FCT through project SCALE-QLT, DOI: 10.54499/2024.16192.PEX.
RP acknowledges funding from the Deutsche Forschungsgemeinschaft (DFG) under the Priority Programme SPP 2514.
K. {\.Z}. acknowledges funding from the European Union under the ERC Advanced Grant TAtypic (project number 101142236).
\end{acknowledgments}
\section*{Author Contributions}
NS conceived the analytical calculations, performed some numerical simulations, and wrote the manuscript. RP carried out the majority of the numerical simulations. PR conceived the original idea, supervised the project, and reviewed the manuscript. SD and KŻ participated in discussions of the project and reviewed the manuscript. All authors discussed the results and approved the final manuscript. Large language models were used for language polishing and coding assistance. All authors reviewed the outputs and take full responsibility for the content of the manuscript.

\section*{Data Availability}
The numerical data and analysis scripts used to generate the figures are available in the public repository \url{https://github.com/nadir265/rqc_error_accumulation_diluted_circuits}.

\onecolumngrid
\appendix
\section{Powers of fidelity}
\label{app:powers_fid}
Let us consider a generic case where we compute the powers of the fidelity $\mathcal{F}^k$ between the output state of an ideal-faultless circuit $\ket{\Psi}$
\begin{equation}
    \ket{\Psi}=U_TU_{T-1},\ldots,U_1\ket{\psi_0},
    \nonumber
\end{equation}
and a noisy output given by a Kraus map
\begin{equation}
\rho=\Phi(\Phi(\ldots\Phi(\ket{\psi_0}\bra{\psi_0})))=\sum_{\mu_1,\ldots,\mu_T=0}^{r+1}\overset{T}{\overleftarrow{\prod_{\tau=1}}}\hat{K}_{\mu_\tau}\ket{\psi_0}\bra{\psi_0}\overrightarrow{\prod_{\tau=1}}\hat{K}^\dagger_{\mu_\tau},
\nonumber
\end{equation}
where we implement $T$ times the quantum channel $\Phi(\rho_\tau)$. For now we are not doing further assumptions with respect the connectivity and the possible further decomposition of each layer $U_\tau$ or $K_{\mu_\tau}$.
Then we can write
\begin{equation}
\begin{split}
\mathcal{F}^k&=\abs{\mel{\Psi}{\rho}{\Psi}}^k\\
&=\left[\bra{\psi_{0}}\left(\overrightarrow{\prod_{\tau=1}}U_{\tau}^{\dagger}\right)\left(\sum_{\mu_1,\ldots,\mu_T=0}^{r+1}\left(\overset{T}{\overleftarrow{\prod_{\tau=1}}}\hat{K}_{\mu_\tau}\right)\ket{\psi_0}\bra{\psi_0}\left(\overrightarrow{\prod_{\tau=1}}\hat{K}^\dagger_{\mu_\tau}\right)\right)\left(\overset{T}{\overleftarrow{\prod_{\tau=1}}}U_{\tau}\right)\ket{\psi_0}\right]^k,
\end{split}
\end{equation}
where arrows above product symbols indicate the stacking order of the gates. To make the above expression more tractable, we shall introduced closure relations $\sum_{\bm{n}}\ketbra{\bm{n}}{\bm{n}}=\mathds{1}$ with $\bm{m}=m_1,\ldots, m_L,\  m_i=0,1.$ and 
use the vectorized notation where density matrices are treated as kets in a duplicated Hilbert space
$$\ket{\psi^{T}} =\bra{\psi}^{T}, \quad \kket{\psi\phi}  =\ket{\psi}\otimes\ket{\phi^{T}}.$$
\begin{equation}
\begin{split}
\mathcal{F}^k&=\abs{\mel{\Psi}{\rho}{\Psi}}^k=\\
&=\sum_{\substack{\bm{m_1},\ldots,\bm{m_k}\\\bm{n_1},\ldots,\bm{n_k}}}\bra{\bm{n_1}\psi_0\bm{m_1}\psi_0\hdots\bm{n_k}\psi_0\bm{m_k}\psi_0}\\
&\hspace{2cm} \times \sum_{\substack{\bm{\mu}^1,\hdots\bm{\mu}^k\\ \bm{\mu}^k=\mu^k_1,\ldots,\mu^k_T}}
\Biggl[\left(\overleftarrow{\prod_{\tau_1=1}}U_{\tau_1}\right)\otimes\left(\overrightarrow{\prod_{\tau=1}}\hat{K}^\dagger_{\mu^1_\tau}\right)\otimes\left(\overset{T}{\overleftarrow{\prod_{\tau=1}}}\hat{K}_{\mu^1_\tau}\right)\otimes\left(\overrightarrow{\prod_{\tau_1=1}}U_{\tau_1}^{\dagger}\right)\Biggr]^{\otimes k}\\
&\hspace{8cm}\times
\ket{\psi_0 \bm{n_1}\psi_0\bm{m_1}\hdots\psi_0 \bm{n_k}\psi_0\bm{m_k}}\\
&=\frac{1}{d}\bbra{\bm{e}_{2k}}\overset{T}{\overleftarrow{\prod_{\tau=1}}}\left[\bigotimes_{i=1}^k\left(U_{\tau}\otimes\left(\sum_{\mu^i_\tau=0}^r\hat{K}^*_{\mu^i_\tau}\otimes\hat{K}_{\mu^i_\tau}\right)\otimes U_{\tau}^*\right)\right]\kket{\boldsymbol{\perp}_{4k}}\ ,
\end{split}
\label{eq:sm_powerF}
\end{equation}
where we have further averaged over the computational basis and defined the states
\begin{eqnarray}
\label{eq:sm_stateperp}
        &\kket{\boldsymbol{\perp}_{4k}}\equiv  \sum_{\bm{l}}\underbrace{\kket{\bm{l}\ldots,\bm{l}}}_{4k}\\
        &\kket{\bm{e}_{2k}}\equiv  \sum_{\substack{\bm{m_1},\ldots,\bm{m_k}\\\bm{n_1},\ldots,\bm{n_k}}}\kket{\bm{n_1}\bm{n_1}\bm{m_1}\bm{m_1}\hdots\bm{n_k}\bm{n_k}\bm{m_k}\bm{m_k}}\ ,
        \label{eq:sm_stateid}
\end{eqnarray}
where the last one corresponds to the identity operation belonging to the symmetric group of $k$ copies. 

It is convenient to explicitly isolate the unitary evolution by defining
\begin{equation}
    \begin{split}
        \mathcal{U}_\tau&\equiv U_{\tau}\otimes U^*_\tau\otimes U_\tau\otimes U_{\tau}^*\ \\\mathcal{K}_{\mu_\tau}&\equiv \mathds{1}\otimes \hat{K}^*_{\mu_\tau}\otimes \hat{K}_{\mu_\tau}\otimes \mathds{1}\\
        \mathcal{W}_\tau&\equiv \overset{\tau}{\overleftarrow{\prod_{\tau'=1}}}\left(\mathds{1}\otimes U^*_{\tau'}\otimes U_{\tau'}\otimes\mathds{1}\right)\ ,
    \label{eq:sm_defcaligraphs}
    \end{split}
\end{equation}
allowing us to write
\begin{equation}
    \mathcal{F}^k=\frac{1}{d}\bbra{\bm{e}_{2k}}\Bigl[\bigotimes_{i=1}^k\overset{T}{\overleftarrow{\prod_{\tau=1}}}\mathcal{U}_\tau\Bigr]\overset{T}{\overleftarrow{\prod_{\tau=1}}}\  \Bigl[\bigotimes_{i=1}^k \sum_{\mu_\tau=0}^{r}\mathcal{W}^\dagger_\tau \mathcal{K_{\mu_\tau}}\mathcal{W}_{\tau-1}\Bigr]\kket{\boldsymbol{\perp}_{4k}}.
\end{equation}
It is easy to see with Eqs. \eqref{eq:sm_stateid} and \eqref{eq:sm_defcaligraphs} that
\begin{equation}
    \bbra{\bm{e}_{2k}}\Bigl[\bigotimes_{i=1}^k\overset{T}{\overleftarrow{\prod_{\tau=1}}}\mathcal{U}_\tau\Bigr]=\bbra{\bm{e}_{2k}},
\end{equation}
yielding the final expression 
\begin{equation}
    \mathcal{F}^k=\frac{1}{d}\bbra{\bm{e}_{2k}}\overset{T}{\overleftarrow{\prod_{\tau=1}}}\  \Bigl[\bigotimes_{i=1}^k \sum_{\mu_\tau=0}^{r}\mathcal{W}^\dagger_\tau \mathcal{K_{\mu_\tau}}\mathcal{W}_{\tau-1}\Bigr]\kket{\boldsymbol{\perp}_{4k}}.
    \label{eq:sm_fidalternative}
\end{equation}

Finally, notice that if we particularize for $k=1$ we obtain
\begin{equation}
    \begin{split}
        \mathcal{F}=\frac{1}{d}\bbra{\bm{e}_{2}}\overset{T}{\overleftarrow{\prod_{\tau=1}}}\left[U_{\tau}\otimes\left(\sum_{\mu^i_\tau=0}^r\hat{K}^*_{\mu_\tau}\otimes\hat{K}_{\mu_\tau}\right)\otimes U_{\tau}^*\right]\kket{\boldsymbol{\perp}_{4}}=\frac{1}{d}\bbra{\bm{e}_{2}}\overset{T}{\overleftarrow{\prod_{\tau=1}}}\  \Bigl[\sum_{\mu_\tau=0}^{r}\mathcal{W}^\dagger_\tau \mathcal{K_{\mu_\tau}}\mathcal{W}_{\tau-1}\Bigr]\kket{\boldsymbol{\perp}_{4}}.
    \end{split}
    \label{eq:sm_fidfromgenerick}
\end{equation}
The first expression corresponds to Eq.\eqref{eq:fidelity_posed} of the main text by identifying $\kket{\boldsymbol{\perp}_{4}}\equiv\kket{\pthree}$, $\kket{\bm{e}_{2}}\equiv\kket{\ptwo}$ and suitable permuting the four copies.

\section{Higher moments of fidelity with (all-to-all) diluted unitaries}
\label{app:highermoments}
Let us particularize the expression for the all-to-all diluted unitaries model, in which our Kraus operators take the form:

\begin{equation}
\begin{split}
    \hat{K}_0&=\sqrt{1-\kappa}\ \tilde{U}\ \text{ verifying } \quad\tilde{U}^\dagger\tilde{U}=\mathds{1}\\
\hat{K}_{\mu\neq0}&=\sqrt{\kappa}\ K_\mu\ \text{ verifying } \quad \sum_{\mu=1}^{r}K^\dagger_\mu K_\mu=\mathds{1}
\end{split}
\end{equation}
for $\kappa\in[0,1]$, and where $\tilde{U}_\tau$ corresponds to the faulty unitary realization of $U_\tau$. 
Then Eq. \eqref{eq:sm_fidalternative} takes the form:

\begin{equation}
        \mathcal{F}^k=\frac{1}{d}\bbra{\bm{e}_{2k}}\overset{T}{\overleftarrow{\prod_{\tau=1}}}\  \Bigl[\bigotimes_{i=1}^k (1-\kappa)\mathcal{W}^\dagger_\tau \tilde{\mathcal{U}}_\tau \mathcal{W}_{\tau-1}+\kappa\sum_{\mu_\tau=1}^{r}\mathcal{W}^\dagger_\tau \mathcal{K_{\mu_\tau}}\mathcal{W}_{\tau-1}\Bigr]\kket{\boldsymbol{\perp}_{4k}},
\end{equation}
with \begin{equation}
   \tilde{\mathcal{U}}_{\tau}\equiv \mathds{1}\otimes \tilde{U}^*_{\tau}\otimes \tilde{U}_{\tau}\otimes\mathds{1}. 
\end{equation}
For simplicity, let us assume that the only source of error is the dissipation. Hence, using Eq. \eqref{eq:sm_defcaligraphs} we find:
\begin{equation}
        \mathcal{F}^k=\frac{1}{d}\bbra{\bm{e}_{2k}}\overset{T}{\overleftarrow{\prod_{\tau=1}}}\  \Bigl[\bigotimes_{i=1}^k (1-\kappa)\mathds{1}+\kappa\sum_{\mu_\tau=1}^{r}\mathcal{W}^\dagger_\tau \mathcal{K_{\mu_\tau}}\mathcal{W}_{\tau-1}\Bigr]\kket{\boldsymbol{\perp}_{4k}},
\end{equation}
Since we are considering an all-to-all geometry and each layer is assumed to be independently sampled, the total average can be factorized:
\begin{equation}
        \overline{\mathcal{F}^k}=\frac{1}{d}\bbra{\bm{e}_{2k}} \Biggl(\ \overline{\Bigl[(1-\kappa)\mathds{1}+\kappa\sum_{\mu_\tau=1}^{r}\mathcal{W}^\dagger_\tau \mathcal{K_{\mu_\tau}}\mathcal{W}_{\tau-1}\Bigr]^{\otimes k}}\ \Biggr)^T\kket{\boldsymbol{\perp}_{4k}},
\end{equation}
Moreover, the expression can be further simplified if the Kraus operators are sampled from the Haar measure. In this case, the left right invariance allows to suppress the unitary strings in each of the terms of the expansion of the tensor product:
\begin{equation}
\overline{\left(\sum_{\mu^i_\tau=1}^r\mathcal{W}^\dagger_\tau\mathcal{K}_{\mu^i_\tau}\mathcal{W}_{\tau-1}\right)^{\otimes i}}=\overline{\left(\sum_{\mu^i_\tau=1}^r\mathcal{K}_{\mu^i_\tau}\right)^{\otimes i} }.
\end{equation}
Hence we arrive to the final expression:
\begin{equation}
        \overline{\mathcal{F}^k}=\frac{1}{d}\bbra{\bm{e}_{2k}} \Biggl(\ \overline{\Bigl[(1-\kappa)\mathds{1}_{4d}+\kappa\sum_{\mu_\tau=1}^{r}\mathcal{K_{\mu_\tau}}\Bigr]^{\otimes k}}\ \Biggr)^T\kket{\boldsymbol{\perp}_{4k}},
\end{equation}

In order to proceed, we make the simplifying choice $T=1$. Since we aim to show that the higher moment concentrates, concentration in the shallowest circuit already supports concentration at larger depths. In this situation, we can harness the isotropic nature of the states Eqs.~\eqref{eq:sm_stateperp} and~\eqref{eq:sm_stateid}
\begin{equation}
    \overline{\mathcal{F}^k}=\frac{1}{d}\left(\sum_{i=0}^{k}\binom{k}{i}(1-\kappa)^{k-i}\kappa^i\bbra{\bm{e}_{2k}}\left(\mathds{1}_{4d}^{\otimes k-i}\otimes\overline{\sum_{\mu^i_\tau=1}^r\mathcal{K}_{\mu^i_\tau}^{\otimes i}}\right)\kket{\boldsymbol{\perp}_{4k}}\right)
\end{equation}
Again, exploiting the same property, we can pack together all the identities taking into account Eq.\eqref{eq:sm_defcaligraphs}
\begin{equation}
    \overline{\mathcal{F}}^k=\frac{1}{d}\left(\sum_{i=0}^{k}\binom{k}{i}(1-\kappa)^{k-i}\kappa^i\bbra{\bm{e}_{2k}}\left(\mathds{1}_{2d}^{\otimes 2k-i}\otimes\overline{\sum_{\mu^i}^r\left(K_{\mu^i}\otimes K^*_{\mu^i}\right)^{\otimes i}}\right)\kket{\boldsymbol{\perp}_{4k}}\right)
    \label{eq:sm_fidbeforeWeingartne}
\end{equation}

Now we focus on computing the average of copies of the sum of Kraus operators. Since each of the Kraus is sampled from the Haar measure, when averaging only matters the number of identical copies $\mathcal{K}_{\mu^i}^{\otimes i}\equiv\left(K_{\mu^i}\otimes K^*_{\mu^i}\right)^{\otimes i}$. This poses a crucial simplification, since we only have to count the total number of different cases that sum a total of $r^i$ terms. To gain intuition, consider $i=2,3$. There are two options:
$$\overline{\left(\sum_{\mu^i=1}^r\mathcal{K}_{\mu^i}\right)^{\otimes 2} } = r\  \overline{\mathcal{K}\otimes \mathcal{K}}+ r(r-1)\ \overline{\mathcal{K}}\otimes \overline{\mathcal{K}}$$
$$\overline{\left(\sum_{\mu^i=1}^r\mathcal{K}_{\mu^i}\right)^{\otimes 3} } = r\  \overline{\mathcal{K}\otimes \mathcal{K}\otimes\mathcal{K}}+ 3 r(r-1)\  \overline{\mathcal{K}\otimes\mathcal{K}}\otimes \overline{\mathcal{K}}+r(r-1)(r-2)\ \overline{\mathcal{K}}\otimes \overline{\mathcal{K}}\otimes  \overline{\mathcal{K}}$$

By induction, it can be seen then that a general expression can be written in terms of all the partitions of the integer $\mathcal{P}(i)$. We denote the $p-$th partition $\bm{\lambda}^p[i]\vdash \ i$ with elements $\bm{\lambda}^p[i]=(\lambda^p_1,\hdots,\lambda^p_{\ell(\bm{\lambda}^p)})$. For example, $\bm{\lambda}^1\ \vdash 3 =(3)$, $\bm{\lambda}^2\ \vdash 3 =(2,1)$ and $\bm{\lambda}^3\ \vdash 3 =(1,1,1)$. We call $\ell(\bm{\lambda}^p[i])$ the length of the partition.
\begin{equation}
\begin{split}
\overline{\left(\sum_{\mu^i=1}^r\mathcal{K}_{\mu^i}\right)^{\otimes i} }&=\sum_{p=1}^{\mathcal{P}(i)} \chi(\bm{\lambda}^p[i])\overline{\mathcal{K}^{\otimes \lambda^p_1}}\otimes\overline{\mathcal{K}^{\otimes \lambda^p_2}}\hdots\otimes\overline{\mathcal{K}^{\otimes \lambda^p_{\ell(\bm{\lambda}^p)}}}\\
     \chi(\bm{\lambda}^p[i])&=\frac{1}{\prod_a^{\ell}\text{multiplicity}(\lambda^p_a)!}\binom{i}{\lambda^p_1\hdots \lambda^p_\ell} \frac{r!}{(r-\ell(\bm{\lambda}^p[i]))!}
\end{split}
\label{eq:sm_chipartition_i}
\end{equation}

 This allows to write the multinomial $i-$th tensor product of $r$ elements as a sum over the partitions $\mathcal{P}(i)$, where $\mathcal{P}(x)$ is the integer partition function of $x$. For each partition of $i$ $\bm{\lambda}^p[i]$ we obtained a coefficient $\chi(\bm{\lambda}^p[i])$ given in Eq.\eqref{eq:sm_chipartition_i}.
Now we can compute each of the above averages $\overline{\mathcal{K}^{\otimes{\lambda^p_a}}}$ via Weingarten calculus~\cite{collinsWeingartenCalculus2022}, yielding a result made of combinations of elements of the symmetric group $S_{\lambda^p_a}$. Making it explicit yields
\begin{equation}
\overline{\mathcal{K}^{\otimes{\lambda_a^p}}}=\overline{\left(K^*\otimes K\otimes \right)^{\otimes \lambda_a^p}}=\sum_{\pi,\tau\in S_{\lambda_a^p}}\Omega(\pi\tau^{-1},dr)\kket{\pi}\bbra{\tau},
\end{equation}
where $\Omega(\pi\tau^{-1},dr)$ are usually called Weingarten functions and
\begin{equation}
\kket{\pi}\bbra{\tau}\equiv \sum_{\substack{\bm{m_in_i}\\i= 1...l}} \kket{\bm{m}_1\bm{m}_{\pi(1)}\ldots\bm{m}_l\bm{m}_{\pi(a)}}\bbra{\bm{n}_1\bm{n}_{\tau(1)}\ldots\bm{n}_l\bm{n}_{\tau(a)}}
\end{equation}
The number of different Weingarten functions corresponds to the number of different classes of $S_{\lambda^p_a}$ which in turn corresponds with the number of partitions of each integer $\lambda^p_a$. Hence we can write Eq.~\ref{eq:sm_fidbeforeWeingartne}
\begin{equation}
\begin{split}
    \overline{\mathcal{F}^k}&=\frac{1}{d}\left(\sum_{i=0}^{k}\binom{k}{i}(1-\kappa)^{k-i}\kappa^i\sum_{p=1}^{\mathcal{P}(i)} \chi(\bm{\lambda}^p[i])\bbra{\bm{e}_{2k}}\left(\mathds{1}_{2d}^{\otimes 2k-i}\otimes\overline{\mathcal{K}^{\otimes \lambda^p_1}}\otimes\overline{\mathcal{K}^{\otimes \lambda^p_2}}\hdots\otimes\overline{\mathcal{K}^{\otimes\lambda^p_{\ell(\bm{\lambda}^p)}}}\right)\kket{\boldsymbol{\perp}_{4k}}\right)\\
    &=\frac{1}{d}\left(\sum_{i=0}^{k}\binom{k}{i}(1-\kappa)^{k-i}\kappa^i\sum_{p=1}^{\mathcal{P}(i)}\chi(\bm{\lambda}^p[i]) \sum_{\pi_1,\tau_1\in S_{\lambda^p_1}}\dots\sum_{\pi_{\ell},\tau_{\ell}\in S_{\lambda^p_{\ell(\lambda^p)}}}\right.\\
    &\hspace{3cm}\left.\prod_{a=1}^{\ell(\bm{\lambda}^p)} \Omega\Bigl(\pi_{\lambda^p_a}\tau_{\lambda^p_a}^{-1},dr\Bigr)
    \bbra{\bm{e}_{2k}}\left(\mathds{1}_{2d}^{\otimes 2k-i}\otimes \kket{\pi_{\lambda^p_1},\dots,\pi_{\lambda^p_\ell}}\bbra{\tau_{\lambda^p_1},\dots,\tau_{\lambda^p_\ell}}\right)\kket{\boldsymbol{\perp}_{4k}}\right)
\end{split}
\end{equation}
Crucially, the inner product is constant:
\begin{equation}
    \bbra{\bm{e}_{2k}}\left(\mathds{1}_{2d}^{\otimes 2k-i}\otimes \kket{\pi_{\lambda^p_1},\dots,\pi_{\lambda^p_\ell}}\bbra{\tau_{\lambda^p_1},\dots,\tau_{\lambda^p_\ell}}\right)\kket{\boldsymbol{\perp}_{4k}}=d,
\end{equation}
so we obtain
\begin{equation}
    \overline{\mathcal{F}^k}=\sum_{i=0}^{k}\binom{k}{i}(1-\kappa)^{k-i}\kappa^i\sum_{p=1}^{\mathcal{P}(i)}\chi(\bm{\lambda}^p[i]) \sum_{\pi_1,\tau_1\in S_{\lambda^p_1}}\dots\sum_{\pi_{\ell},\tau_{\ell}\in S_{\lambda^p_{\ell(\lambda^p)}}}\prod_{a=1}^{\ell(\bm{\lambda}^p)} \Omega\Bigl(\pi_{\lambda^p_a}\tau_{\lambda^p_a}^{-1},dr\Bigr)
\end{equation}

Consider now the case $k=1$; then $\mathcal{P}(1)=1$ and we obtain:
\begin{equation}
    \overline{\mathcal{F}}=(1-\kappa)+\kappa\sum_{p=1}^1\chi(\bm{\lambda}^1[1]) \sum_{\pi_1,\tau_1\in S_{\lambda^p_1}} \Omega\Bigl(\pi_{\lambda^1_1}\tau_{\lambda^1_1}^{-1},dr\Bigr)=(1-\kappa)+\kappa \frac{r}{dr}=1-\kappa(1-\frac{1}{d}).
    \label{eq:sm_fid1layer}
\end{equation}

Let us move to $k=2$, then $\mathcal{P}(2)=2$, $\bm{\lambda}^1[1]=(1),\bm{\lambda}^1[2]=(2),\bm{\lambda}^2[2]=(1,1)$.
\begin{equation} 
\begin{split}
    \overline{\mathcal{F}^2}&=\binom{2}{0}(1-\kappa)^2+\binom{2}{1}(1-\kappa)\kappa\chi(\bm{\lambda}^1[1])\Omega(ee^{-1},dr)\nonumber \\
&\quad\quad+\binom{2}{2}\kappa^2 \left(\chi(\bm{\lambda}^1[2])\sum_{\pi,\tau\in S_2}\Omega(\pi\tau^{-1},dr)+\chi(\bm{\lambda}^2[2])\Omega^2(ee^{-1},dr)\right)\nonumber \\
    &=\overline{\mathcal{F}}^2+\kappa^2\left(-\frac{1}{d^2r}+\frac{2r}{d^2r^2-1}-\frac{2}{d(d^2r^2-1)}\right),
\end{split}
\end{equation}
where in the last equality we have used Eq.~\eqref{eq:sm_fid1layer}. Then it is clear that the leading correction is
\begin{equation}
    |\overline{\mathcal{F}^2}-\overline{\mathcal{F}}^2|\sim\frac{\kappa^2}{d^2r}.
\end{equation}
Compute exactly higher moments becomes a difficult task since the amount of different term in the sum grows polynomially with the square of the number of elements of the symmetric group. However, we may neglect all the Weingarten functions except the one associated to the identity class, which is the biggest. As second approximation, we shall assume that they are equal to the one of the $S_1$ group $\Omega(ee^{-1},dr)=\frac{1}{dr}$.
\begin{equation}
\begin{split}
\overline{\mathcal{F}^k}&=\sum_{i=0}^{k}\binom{k}{i}(1-\kappa)^{k-i}\kappa^i\sum_{p=1}^{\mathcal{P}(i)}\chi(\bm{\lambda}^p[i]) \prod_{a=1}^{\ell(\bm{\lambda}^p)} \lambda_a^p! \Bigl(\frac{1}{dr}\Bigr)^i\\&=\sum_{i=0}^{k}\frac{k!}{(k-i)!}(1-\kappa)^{k-i}\kappa^i\Bigl(\frac{1}{dr}\Bigr)^i\sum_{p=1}^{\mathcal{P}(i)}\frac{1}{\prod_a^{\ell}\text{multiplicity}(\lambda^p_a)!}\frac{r!}{(r-\ell(\bm{\lambda}^p[i]))!}
\end{split}
\end{equation}
Observe that the addend
$\frac{1}{\prod_a^{\ell}\text{multiplicity}(\lambda^p_a)!}\frac{r!}{(r-\ell(\bm{\lambda}^p[i]))!}$
counts the number of ways of distributing $i$ indistinguishable balls into
$r$ distinguishable boxes with occupation pattern $\bm{\lambda}^p$: the factor
$r!/(r-\ell)!$ assigns the $\ell(\bm{\lambda}^p)$ parts to $\ell$ ordered
distinct boxes, while dividing by the multiplicities removes the spurious
ordering of equal parts. Summing over all partitions $p$ therefore amounts to
summing over all possible occupation patterns, which exhausts the whole set of
distributions,
\begin{equation}
    \sum_{p=1}^{\mathcal{P}(i)}
    \frac{1}{\prod_a^{\ell}\text{multiplicity}(\lambda^p_a)!}
    \frac{r!}{(r-\ell(\bm{\lambda}^p[i]))!}
    =\binom{r+i-1}{i}=\frac{(r)_i}{i!}\ ,
\end{equation}
where $(r)_i=r(r+1)\cdots(r+i-1)$ is the Pochhammer symbol. 
Thus, we have
\begin{equation}
\overline{\mathcal{F}^k}=\sum_{i=0}^{k}\binom{k}{i}(1-\kappa)^{k-i}\kappa^i\Bigl(\frac{1}{d}\Bigr)^i\frac{(r)_i}{r^i}, \quad \text{and} \quad \overline{\mathcal{F}}^k= \sum_{i=0}^{k}\binom{k}{i}(1-\kappa)^{k-i}\kappa^i\Bigl(\frac{1}{d}\Bigr)^i,
\end{equation}
and therefore
\begin{equation}
    \Delta_k=\sum_{i=0}^{k}\binom{k}{i}(1-\kappa)^{k-i}\kappa^i\frac{1}{d^i}\Bigl(\frac{(r)_i}{r^i}-1\Bigr)=\frac{1}{d^2r}\binom{k}{2}(1-\kappa)^{k-2}\kappa^2+\sum_{i=3}^{k}\binom{k}{i}(1-\kappa)^{k-i}\kappa^i\frac{1}{d^i}\Biggl(\frac{(r)_i}{r^i}-1\Biggr).
\end{equation}
The terms $i=0$ and $i=1$ cancel identically, since $(r)_0=1$ and $(r)_1=r$:
a single error cannot collide with another one and the average factorizes
exactly. The leading contribution is therefore $O(\kappa^2)$, the remainder
being $O(\kappa^3/d^3)$.

To finish this appendix, observe that we are interested in circuits of depth
$T$. In the simplest scenario, assuming an independent Haar unitary at each
layer, the moments factorize across layers $\overline{\mathcal{F}_T^k}=\overline{\mathcal{F}^k}^T$ and $\overline{\mathcal{F}_T}^k=\overline{\mathcal{F}}^{kT}$, so linearizing the rest yields
\begin{equation}
    \Delta_k^{(T)}\simeq\frac{T}{d^2r}\binom{k}{2}(1-\kappa)^{Tk-2}\kappa^2\ ,
\end{equation}
which is the expression appearing in the main text.

\section{Universal form of fidelity}
\label{app:universalfid}
On this Appendix we shall derive the form of the average fidelity given in the main text Eq. \eqref{eq:universal_fid} associated with the RQC given in Fig. \ref{fig:layer_errors}(b). We start with the formal expression of the fidelity for a dissipative random quantum circuit of the form depicted in Fig. \ref{fig:layer_errors}(b) which is obtained by particularizing Eq.\eqref{eq:sm_fidfromgenerick}:
\begin{equation}
  \mathcal{F}=\frac{1}{d}\bbra{\pone}
    \overset{T}{\overleftarrow{\prod_{\tau=1}}}\Bigl[ 
  \bm{\mathcal{V}}_\tau\ \bm{\Pi}_\tau\Bigr]\kket{\pthree},  
\end{equation}
where we define
\begin{equation}
\begin{split}
\bm{\mathcal{V}}&\equiv\bigotensor_{m=1}^{L/2}\Bigl((1-\kappa)\bm{u}_{2m-1,2m}
   \ \ +\kappa\sum^r_{\mu_{2m-1,2m}=1} \bm{k}_{\mu_{2m-1,2m}}\Bigr)\\&\qquad \text{with}\quad \begin{cases}
       \bm{u}_{m,m'}&\equiv u_{m,m'}\otimes u^*_{m,m'}\otimes u_{m,m'}\otimes u^*_{m,m'}\\
        \bm{k}_{\mu_{mm'}}&\equiv k_{\mu_{m,m'}}\otimes k^*_{\mu_{m,m'}}\otimes u_{m,m'}\otimes u^*_{m,m'}.
   \end{cases} \quad \text{and}\\
\bm{\Pi}_\tau&\equiv\Pi_\tau\otimes\Pi_\tau\otimes\Pi_\tau\otimes\Pi_\tau\ .
\end{split}
\label{eq:sm_defVPi}
\end{equation}
As discussed in the main text, we make a crucial simplification consisting in replacing the random all-to-all permutations by a random all-to-all unitary matrix $\Pi_\tau\to R_\tau$ with $R\in\text{CUE}(d)$. Hence we find that the average of the simplified version of the fidelity is 
\begin{equation}
    \overline{\mathcal{F}_\text{simp}}=\frac{1}{d}\bbra{\pone}
    \Bigl(\overline{ 
  \bm{\mathcal{V}}}\ \overline{\ \bm{\mathcal{R}}}\Bigr)^T\kket{\pthree}=\frac{1}{d}\left(\bbra{\pone}\overline{\bm{\mathcal{V}}}\right)\ \left(\overline{\bm{\mathcal{R}}}\ \overline{\bm{\mathcal{V}}}\right)^{T-1}\ \left(\overline{\bm{\mathcal{R}}}\kket{\pthree}\right),
  \label{eq_sm_3partsfid}
\end{equation}
where in the last equality, we have split the computation in three parts for future convenience.

 The averages of copies of unitaries sampled uniformly have been widely studied and computed by means of Weingarten calculus~\cite{collinsWeingartenCalculus2022}. 
 In particular, the average of one single copy is
 \begin{equation}
     \overline{R\otimes R^*}=\frac{1}{d}\kket{\bm{+^{12}}}\bbra{\bm{+^{12}}}=\frac{1}{d}\ \ketub\braub
     \label{eq:sm_haar1copy}
 \end{equation}
 
 over two copies (and its conjugates) is given in terms of elements of two elements of the $S_2$ symmetric group. This can easily be seen in terms of the diagrams showing the trivial (identity) and non-trivial (swap) ways of connecting a unitary with a conjugate unitary:
\begin{equation}
\begin{split}
\overline{\bm{\mathcal{R}}}=\overline{(R\otimes R^{*})^{\otimes 2}}&=\frac{1}{d^2-1}\Biggl(\kket{\ptwo}\bbra{\ptwo}+\kket{\pone}\bbra{\pone}\\
&\hspace{2cm}-\frac{1}{d}\Bigl(\kket{\pone}\bbra{\ptwo}+\kket{\ptwo}\bbra{\pone}\Bigr)\Biggr)\\
      &=\frac{1}{d^2-1}\Biggl( \ \ketu \brau \ + \ \ketv \brav \ - \ \frac{1}{d}\Biggl(\ \ketu \brav \ +\ \ketv\brau \ \Biggr)\ \Biggr) \ ,  
\end{split}
    \label{eq:sm_haar2copies}
\end{equation}

Observe that the states $\kket{\pone}$, $\kket{\ptwo}$ and $\kket{\pthree}$ are not orthogonal:
\begin{eqnarray}
     \bbrakket{\ptwo}{\ptwo}= \brau\hspace{-0.5pt}\ketu \ = d^2 = \ \brav\hspace{-0.5pt}\ketv \ =  \bbrakket{\pone}{\pone} \nonumber\\
    \bbrakket{\ptwo}{\pone}= \brau\hspace{-0.5pt}\ketv \ = d = \ \brau\hspace{-0.5pt}\ketv \ =  \bbrakket{\pone}{\ptwo} \\
    \bbrakket{\pthree}{\pone}= \braz\hspace{-0.5pt}\ketv \ = d = \ \brau\hspace{-0.5pt}\ketz \ =  \bbrakket{\pthree}{\ptwo}.
    \label{eq:sm_innerstates}
\end{eqnarray}
Hence, we find an orthogonal basis via the Gram-Schmidt procedure:
\begin{equation}
\begin{split}
    \kket{\Up} &=\frac{1}{d} \kket{\ptwo}\ \quad \ \kket{\Down}=\frac{1}{\sqrt{d^2-1}}\left(\kket{\pone}-\frac{1}{d}\kket{\ptwo}\right)\\
      \kket{\ptwo}&= d\kket{\Up}\ \quad \ \kket{\pone}=\sqrt{d^2-1}\kket{\Down}+\kket{\Up}.
\end{split}
    \label{eq:sm_spinbasis}
\end{equation}
On this basis, the expression Eq.~\eqref{eq:sm_haar2copies} takes the simpler form
\begin{equation}
\overline{\bm{\mathcal{R}}}=\kket{\Up}\bbra{\Up}+\kket{\Down}\bbra{\Down}. 
    \label{eq:sm_avg_haar_diag}
\end{equation}
\begin{equation}
\begin{split}
    \overline{\mathcal{F}_\text{simp}}&=\frac{1}{d}\left(\bbra{\pone}\overline{\bm{\mathcal{V}}}\right)\ \Bigl(\kket{\Up}\bbra{\Up}\overline{\bm{\mathcal{V}}}+\kket{\Down}\bbra{\Down}\ \overline{\bm{\mathcal{V}}}\Bigr)^{T-1}\ \left(\kket{\Up}\bbrakket{\Up}{\pthree}+\kket{\Down}\bbrakket{\Down}{\pthree}\right)\\
    &=\frac{1}{d}\left(\sqrt{d^2-1} \bbra{\Down}\overline{\bm{\mathcal{V}}}+\bbra{\Up}\overline{\bm{\mathcal{V}}}\right)\ \Bigl(\kket{\Up}\bbra{\Up}\overline{\bm{\mathcal{V}}}+\kket{\Down}\bbra{\Down}\ \overline{\bm{\mathcal{V}}}\Bigr)^{T-1}\ \left(\kket{\Up}+\sqrt{\frac{d-1}{d+1}}\kket{\Down}\right),
    \end{split}
  \label{eq_sm_3partsfid1}
\end{equation}
where in the second equality we have used Eq.~\eqref{eq:sm_spinbasis} and Eq.~\eqref{eq:sm_innerstates}. Hence, we need to evaluate both $\bbra{\Up}\overline{\bm{\mathcal{V}}}$ and $\bbra{\Down}\overline{\bm{\mathcal{V}}}$. To do so, we write all the terms on the same local basis. The easiest starting point is $\kket{\Up}$, since
\begin{align}
    \kket{\Up}&=\bigotimes_{r=1}^{L/2}\kket{\Up_{2m-1,2m}}  \label{eq:sm_up_newbasis}\\
    \kket{\Down}&=\frac{1}{\sqrt{d^2-1}}\lp\bigotimes_{m=1}^{L/2}\lp\kket{\Up_{2m-1,2m}} + \sqrt{15}\kket{\Down_{2m-1,2m}}\rp-\bigotimes_{m=1}^{L/2}\kket{\Up_{2m-1,2m}}\rp,
    \label{eq:sm_down_newbasis}
\end{align}
where in the second equality we have inverted Eq. \eqref{eq:sm_spinbasis} and particularized for $d=4$. Taking into account the definition of $\bm{\mathcal{V}}$ given in Eq.~\eqref{eq:sm_defVPi}, we have:
\begin{equation}
\begin{split}
\overline{\bm{\mathcal{V}}}&=\overline{\bigotensor_{m=1}^{L/2}\Bigl((1-\kappa)\bm{u}_{2m-1,2m}+\kappa\sum^r_{\mu_{2m-1,2m}=1} \bm{k}_{\mu_{2m-1,2m}}\Bigr)}\\
&=\bigotensor_{m=1}^{L/2}\Bigl((1-\kappa)\overline{\bm{u}_{2m-1,2m}}+\kappa\overline{\sum^r_{\mu_{2m-1,2m}=1}\bm{k}_{\mu_{2m-1,2m}}}\Bigr).
\end{split}
\label{eq:sm_avgtensordiluted}
\end{equation}
Putting together Eqs.\eqref{eq:sm_avgtensordiluted}\eqref{eq:sm_up_newbasis} and \eqref{eq:sm_down_newbasis} yields
\begin{align}
\bbra{\Up}\overline{\bm{\mathcal{V}}}&=\bigotensor_{m=1}^{L/2}\Bigl((1-\kappa)\bbra{\Up_{2m-1,2m}}\overline{\bm{u}_{2m-1,2m}}+\kappa \sum^r_{\mu_{2m-1,2m}=1}\bbra{\Up_{2m-1,2m}}\bm{k}_{\mu_{2m-1,2m}}\Bigr)\nonumber \\
&=\bigotensor_{m=1}^{L/2}\Biggl[\ \frac{(1-\kappa)}{4} \text{Avg}\ \leftparenthesis\brau \fourops{u_{2m-1,2m}}{u^*_{2m-1,2m}}{u_{2m-1,2m}}{u^*_{2m-1,2m}} \ \rightparenthesis \ + \frac{\kappa}{4} \sum^r_{\mu_{2m-1,2m}=1} \text{Avg}\ \leftparenthesis \brau \fourops{k_{\mu_{2m-1,2m}}}{k^*_{\mu_{2m-1,2m}}}{u_{2m-1,2m}}{u^*_{2m-1,2m}}\rightparenthesis\ \Biggr]\\\nonumber
&=\bigotensor_{m=1}^{L/2}\frac{1}{4}\bbra{\ptwo_{2m-1,2m}} = \bbra{\Up},
\end{align}
where we have used in the second line the unitarity $U^\dagger U=\mathds{1}$ and also the constraint $\sum_{\mu=1}^{r}\hat{K}^\dagger_\mu \hat{K}_\mu=\mathds{1}$. In the same way, 
\begin{equation}
\begin{split}
\bbra{\Down}\overline{\bm{\mathcal{V}}}&=\frac{1}{\sqrt{d^2-1}}\Bigl(\bigotensor_{m=1}^{L/2}\bbra{\pone_{2m-1,2m}}-\bbra{\Up}\Bigr)\bigotensor_{m=1}^{L/2}\Bigl((1-\kappa)\overline{\bm{u}_{2m-1,2m}}+\kappa\overline{\sum^r_{\mu_{2m-1,2m}=1}\bm{k}_{\mu_{2m-1,2m}}}\Bigr)\\
&=\frac{1}{\sqrt{d^2-1}}\bigotensor_{m=1}^{L/2}\leftparenthesis(1-\kappa)\bbra{\pone_{2m-1,2m}} \ \\
&\hspace{2cm} + \kappa \sum^r_{\mu_{2m-1,2m}=1}\brav \twoboxes{\overline{k_{\mu_{2m-1,2m}}\otimes k^*_{\mu_{2m-1,2m}}}}{\overline{u_{2m-1,2m}\otimes u^*_{2m-1,2m}}}\rightparenthesis-\frac{1}{\sqrt{d^2-1}}\bbra{\Up}
\label{eq:sm_downV}
\end{split}
\end{equation}

Thus, we observe that for the majority of the terms, the detailed structure of the two-qubit unitaries and of the Kraus noise does not play any role. This is a manifestation of the right invariance of the Haar measure induced by the substitution of the random permutation matrices by Haar unitaries. Indeed, only one term in Eq.~\eqref{eq:sm_downV} depends on the details of the dissipative quantum circuit. 

Thus, in the main text we consider two possibilities that crucially yield the same result
\begin{enumerate}
    \item The unitary gates are generic and the Kraus operators are arbitrary. Using Eq.\eqref{eq:sm_haar1copy} and particularizing for $d=4$:
    \begin{equation}
        \sum^r_{\mu_{2m-1,2m}=1}\brav \twoboxes{\overline{k_{\mu_{2m-1,2m}}\otimes k^*_{\mu_{2m-1,2m}}}}{u_{2m-1,2m}\otimes u^*_{2m-1,2m}}=\frac{1}{4}\ \brav\upboxbrau{\sum^r_{\mu=1}k_\mu\otimes k^*_\mu}=\frac{1}{4} \brau=\frac{1}{4}\bbra{\ptwo}
        \label{eq:sm_krauscase1}
    \end{equation}
        \item The Kraus operators are generic and the unitaries are arbitrary.  Using again Eq.\eqref{eq:sm_haar1copy} but taking into account that the dimension is $4r$ and that there are $r$ independent averages:
    \begin{equation}
        \sum^r_{\mu_{2m-1,2m}=1}\brav \twoboxes{k_{\mu_{2m-1,2m}}\otimes k^*_{\mu_{2m-1,2m}}}{\overline{u_{2m-1,2m}\otimes u^*_{2m-1,2m}}}=\frac{r}{4r}\ \brav\downboxbrau{u_{2m-1,2m}\otimes u^*_{2m-1,2m}}=\frac{1}{4} \brau=\frac{1}{4}\bbra{\ptwo}
        \label{eq:sm_krauscase2}
    \end{equation}
\end{enumerate}

We therefore conclude that Eq. \eqref{eq:sm_downV} takes the form
\begin{align}
\bbra{\Down}\overline{\bm{\mathcal{V}}}&=\frac{1}{\sqrt{d^2-1}}\Biggl(\bigotensor_{m=1}^{L/2}\Bigl((1-\kappa)\bbra{\pone_{2m-1,2m}} \ + \frac{\kappa}{4} \bbra{\ptwo_{2m-1,2m}} \Bigr)-\bigotensor_{m=1}^{L/2}\frac{1}{4}\bbra{\ptwo_{2m-1,2m}}\Biggr)\nonumber \\
&=\frac{1}{\sqrt{d^2-1}}\Biggl(\bigotensor_{m=1}^{L/2}\Bigl((1-\kappa)\sqrt{15}\bbra{\Down_{2m-1,2m}}+\bbra{\Up_{2m-1,2m}}\Bigr)-\bigotensor_{m=1}^{L/2}\bbra{\Up_{2m-1,2m}}\Biggr)
\label{eq:sm_downV1}
\end{align}

To conclude the calculation, it is convenient to introduce a new basis $\{\kket{m,i}\}$ where $m$ refers to subspace of states with  $m$ down spins $\kket{\Down}$ and $i=1,\dots \binom{L/2}{m}$ labels the state within it. With this notation, 
\begin{equation}
\begin{split}
\bbra{\Up}\overline{\bm{\mathcal{V}}}&=\bbra{\Up} = \bbra{0,1}\\
\bbra{\Down}\overline{\bm{\mathcal{V}}}&=\frac{1}{\sqrt{d^2-1}}\sum_{m=1}^{L/2}(1-\kappa)^m (15)^{m/2}\sum_{i=1}^{\binom{L/2}{m}}\bbra{m,i}\\
\bbra{\Down} &= \frac{1}{\sqrt{d^2-1}}\sum_{m=1}^{L/2}(15)^{m/2}\sum_{i=1}^{\binom{L/2}{m}}\bbra{m,i}
\label{eq:sm_overlapsV_magnonbasis}
\end{split}
\end{equation}

Then it is easy to see that
\begin{align}
&\bbra{\Up}\overline{\bm{\mathcal{V}}}\kket{\Up}=1,\quad \bbra{\Up}\overline{\bm{\mathcal{V}}}\kket{\Down}=\bbra{\Down}\overline{\bm{\mathcal{V}}}\kket{\Up}=0\\
&\bbra{\Down}\overline{\bm{\mathcal{V}}}\kket{\Down}=\frac{1}{d^2-1}\sum_{m,n=1}^{L/2}(1-\kappa)^m 15^{\frac{m+n}{2}}\sum_{i,j=1}^{\binom{L/2}{m}\binom{L/2}{n}}\delta_{m,n}\delta_{i,j}= \frac{(16-15\kappa)^{L/2}-1}{d^2-1}  
\end{align}
Equipped with this we can evaluate the central part in Eq. \eqref{eq_sm_3partsfid1}, since all crossed terms vanish due to the orthogonality of the states with different magnons.
\begin{equation}
\Bigl(\kket{\Up}\bbra{\Up}\overline{\bm{\mathcal{V}}}+\kket{\Down}\bbra{\Down}\overline{\bm{\mathcal{V}}}\Bigr)^{T-1}= \kket{\Up}\bbra{\Up}+\left(\bbra{\Down}\overline{\bm{\mathcal{V}}}\kket{\Down}\right)^{T-2}\kket{\Down}\bbra{\Down}\overline{\bm{\mathcal{V}}} 
\end{equation}
Hence, we obtain the final result:
\begin{equation}
\begin{split}
\overline{\mathcal{F}_\text{simp}}&=\frac{1}{d}\left(\sqrt{d^2-1} \bbra{\Down}\overline{\bm{\mathcal{V}}}+\bbra{\Up}\right)\ \Bigl(\kket{\Up}\bbra{\Up}+\left(\bbra{\Down}\overline{\bm{\mathcal{V}}}\kket{\Down}\right)^{T-2}\kket{\Down}\bbra{\Down}\overline{\bm{\mathcal{V}}} \Bigr)\ \left(\kket{\Up}+\sqrt{\frac{d-1}{d+1}}\kket{\Down}\right),
\\&=\frac{1}{d}+\left(\bbra{\Down}\overline{\bm{\mathcal{V}}}\kket{\Down}\right)^{T}\Bigl(1-\frac{1}{d}\Bigr)\ .
\end{split}
\end{equation}

\section{Fidelity with unitary and dissipative noise}
\label{app:fidunitdis}
The derivation is similar to the one done in App.~\ref{app:universalfid}, but for the benefit of the reader this Appendix will also be self-contained.
We start with the formal expression of the fidelity for a dissipative random quantum circuit of the form depicted in Fig. \ref{fig:layer_errors}(b) which is obtained by particularizing Eq.\eqref{eq:sm_fidfromgenerick}:
\begin{equation}
  \mathcal{F}=\frac{1}{d}\bbra{\pone}
    \overset{T}{\overleftarrow{\prod_{\tau=1}}}\Bigl[ 
  \bm{\mathcal{\tilde{V}}}_\tau\ \bm{\tilde{\Pi}}_\tau\Bigr]\kket{\pthree},  
\end{equation}
where in this case we have
\begin{equation}
\begin{split}
\bm{\mathcal{\tilde{V}}}&\equiv\bigotensor_{m=1}^{L/2}\Bigl((1-\kappa)\bm{\tilde{u}}_{2m-1,2m}
   \ \ +\kappa\sum^r_{\mu_{2m-1,2m}=1} \bm{k}_{\mu_{2m-1,2m}}\Bigr)\\
   &\hspace{1cm}\text{with}\quad \begin{cases}
       \bm{\tilde{u}}_{m,m'}&\equiv \tilde{u}_{m,m'}\otimes \tilde{u}^*_{m,m'}\otimes u_{m,m'}\otimes u^*_{m,m'}\\
        \bm{k}_{\mu_{mm'}}&\equiv k_{\mu_{m,m'}}\otimes k^*_{\mu_{m,m'}}\otimes u_{m,m'}\otimes u^*_{m,m'}.
   \end{cases}\\
\bm{\tilde{\Pi}}_\tau&\equiv\tilde{\Pi}_\tau\otimes\tilde{\Pi}_\tau\otimes\Pi_\tau\otimes\Pi_\tau,
\end{split}
\label{eq:sm_defVPi2}
\end{equation}
where we emphasize the unitary deviation with tilde over all quantities. With the same philosophy than in the previous case, we assume each random permutation is embedded in two independent random all-to-all unitary matrix $\Pi_\tau \to R_{1,\tau}\Pi_\tau R_{2,\tau}$ with $R\in\text{CUE}(d)$. Hence we find that the average of the simplified version of the fidelity is 
\begin{equation}
\begin{split}
    \overline{\mathcal{F}_\text{simp}}&=\frac{1}{d}\bbra{\pone}
    \Bigl(\overline{ 
  \bm{\tilde{\mathcal{V}}}}\ \overline{\bm{\mathcal{R}}\bm{\Pi}\bm{\mathcal{R}}}\Bigr)^T\kket{\pthree}\\
  &=\frac{1}{d}\left(\bbra{\pone}\overline{\bm{\tilde{\mathcal{V}}}}\right)\ \left(\overline{\bm{\mathcal{R}}\bm{\Pi}\bm{\mathcal{R}}}\ \overline{\bm{\tilde{\mathcal{V}}}}\right)^{T-1}\ \left(\overline{\bm{\mathcal{R}}\bm{\Pi}\bm{\mathcal{R}}}\kket{\pthree}\right),
\end{split}
  \label{eq_sm_3partsfidtilde}
\end{equation}

\subsection{\texorpdfstring{Calculation of \boldmath $\mathcal{R}\ \Pi \ \mathcal{R}$}{Calculation of R Pi R}}
Now we shall focus on the average of the faulty permutation contribution. As a first step, we harness the left/right invariance of the Haar measure:
\begin{align}
        \bm{\mathcal{R}}\ \bm{\Pi} \ \bm{\mathcal{R}}&=\lp\overline{R\otimes R^*\otimes R \otimes R^* }\rp \ \lp\overline{\tilde{\Pi}\Pi^T\otimes\tilde{\Pi}\Pi^T \otimes \mathds{1}\otimes \mathds{1}}\rp \ \lp\overline{R\otimes R^*\otimes R \otimes R^* }\rp.\\
&= \frac{1}{(d^2-1)^2}\Biggl[ \ \Biggl( \ \ketu -\ \frac{1}{d} \ \ketv \Biggr) \  \brau\ \ +\ \Biggl(\ \ketv -\frac{1}{d} \ \ketu \Biggr) \  \brav \  \Biggr]\\
&\times \quad\boxfourlegs{\Xi\otimes\Xi}\times \quad\Biggl[ \ \ketu \Biggl(  \brau -\ \frac{1}{d} \ \brav\  \Biggr) \ +\ \ketv \  \Biggl( \brav -\frac{1}{d} \ \brau \ \Biggr)  \  \Biggr],
\end{align}
where $\Xi=\tilde{\Pi}\Pi^T$. Using Eq. \eqref{eq:sm_trboxes} we find that

\begin{eqnarray}
   \label{eq:sm_deltadef}
\brau\hspace{-0.5pt}\boxfourlegs{\Xi\otimes\Xi}\hspace{-0.5pt}\ketu=d \boxtwotrA{\Xi\otimes\Xi}= d^2\quad &,&\quad \brav\hspace{-0.5pt}\boxfourlegs{\Xi\otimes\Xi}\hspace{-0.5pt}\ketv\ = \boxtwotrB{\Xi\otimes\Xi} = \overline{\tr^2{\tilde{\Pi}\Pi^T}} := \delta.\\
\brav\hspace{-0.5pt}\boxfourlegs{\Xi\otimes\Xi}\hspace{-0.5pt}\ketu &=& \brau\hspace{-0.5pt}\boxfourlegs{\Xi\otimes\Xi}\hspace{-0.5pt}\ketv= \boxtwotrA{\Xi\otimes\Xi} =d.\nonumber
\end{eqnarray}
Therefore, we find that 
\begin{equation}
\begin{split}
    \bm{\mathcal{R}}\ \bm{\Pi} \ \bm{\mathcal{R}}&= \frac{1}{(d^2-1)^2}\Biggl\{(d^2-2+\frac{\delta}{d^2})\ketu\brau+(\delta-1)\Biggl(\ketv\brav-\frac{1}{d}\ketu\brav-\frac{1}{d}\ketv\brau\Biggr)\Biggr\}\\
&= \kket{\Up}\bbra{\Up}+\frac{\delta-1}{d^2-1}\kket{\Down}\bbra{\Down},
    \label{eq:sm_avgRPR}
\end{split}
\end{equation}
Where we have used the spin like basis Eq.\eqref{eq:sm_spinbasis}. Hence, we shall write the average fidelity Eq.\eqref{eq_sm_3partsfidtilde} as:
\begin{equation}
\begin{split}
    \overline{\mathcal{F}_\text{simp}}&=\frac{1}{d}\left(\sqrt{d^2-1} \bbra{\Down}\overline{\tilde{\bm{\mathcal{V}}}}+\bbra{\Up}\overline{\tilde{\bm{\mathcal{V}}}}\right)\ \Bigl(\kket{\Up}\bbra{\Up}\overline{\bm{\tilde{\mathcal{V}}}}\\
    &\qquad+\frac{\delta-1}{d^2-1}\kket{\Down}\bbra{\Down}\ \overline{\bm{\tilde{\mathcal{V}}}}\Bigr)^{T-1}\ \left(\kket{\Up}+\frac{\delta-1}{d^2-1}\frac{d-1}{\sqrt{d^2-1}}\kket{\Down}\right),
    \end{split}
  \label{eq_sm_3partsfid2}
\end{equation}
which is instructive to compare with Eq.\eqref{eq_sm_3partsfid1} where there is only dissipative noise.

To conclude, we need to evaluate $\bbra{\Down}\overline{\tilde{\bm{\mathcal{V}}}}$ and $\bbra{\Up}\overline{\tilde{\bm{\mathcal{V}}}}$, where in the same way that Eq.~\eqref{eq:sm_avgtensordiluted}
\begin{equation}
\begin{split}
    \overline{\tilde{\bm{\mathcal{V}}}}&=\overline{\bigotensor_{m=1}^{L/2}\Bigl((1-\kappa)\bm{\tilde{u}}_{2m-1,2m}+\kappa\sum^r_{\mu_{2m-1,2m}=1} \bm{k}_{\mu_{2m-1,2m}}\Bigr)}\\&=\bigotensor_{m=1}^{L/2}\Bigl((1-\kappa)\overline{\bm{\tilde{u}}_{2m-1,2m}}+\kappa\overline{\sum^r_{\mu_{2m-1,2m}=1}\bm{k}_{\mu_{2m-1,2m}}}\Bigr).
\label{eq:sm_avgtensordilutedtilde}
\end{split}
\end{equation}
\subsubsection{\texorpdfstring{Calculation of \boldmath $\overline{\tilde{u}_{2m-1,2m}}$}{Calculation of averaged faulty two-qubit gates}}

As discussed in the main text, each faulty gate is modeled as a random unitary $u_{m,m'}\in\text{CUE}(4)$ perturbed by an unstructured random unitary whose generator belongs to the Gaussian unitary ensemble, $\tilde{u}_{2m-1,2m}=e^{i\alpha h_{2m-1,2m}}u_{2m-1,2m}$, with $h_{2m-1,2m}\in\text{GUE}(4)$ and $\alpha\geq0$.
Therefore, the local average of the 4-copied unitaries can be further decomposed.
\begin{eqnarray}
    \overline{\bm{\tilde{u}_{2m-1,2m}}}&=&\overline{\tilde{u}_{2m-1,2m}\otimes\ \tilde{u}^*_{2m-1,2m}\otimes\ u_{2m-1,2m}\otimes\ u^*_{2m-1,2m}}\\\nonumber
    &=&\left(\overline{e^{i\alpha h_{2m-1,2m}}\otimes\ e^{-i\alpha h_{2m-1,2m}}\otimes\ \mathds{1}\otimes\ \mathds{1}}\right)\lp\overline{u_{2m-1,2m}\otimes\ u^*_{2m-1,2m}\otimes\ u_{2m-1,2m}\otimes\ u^*_{2m-1,2m}}\rp.
    \label{eq:sm_avg_faultygates}
\end{eqnarray}
The rightmost average is given by Eq.\eqref{eq:sm_haar2copies} particularized for $d=4$. Hence, we need to compute $\overline{e^{i\alpha h_{2m-1,2m}}\otimes\ e^{-i\alpha h_{2m-1,2m}}}$ with respect to the GUE measure. 
\begin{equation}
   \overline{e^{i\alpha H}\otimes e^{-i\alpha H^*}}=\int dH e^{-\tr \frac{H^2}{2}}\left(e^{-i\alpha H}\otimes e^{i\alpha H^*}\right),\quad dH=\prod_{i=1}^d d H_{ii}\prod_{i>j}\frac{1}{\sqrt{2}}\text{Re}(H_{ij})\text{Im}({H_{ij}}).
   \label{eq:sm_avggue}
\end{equation}

Crucially, the GUE measure is invariant under unitary transformations, i.e. the unitary matrix that diagonalizes $H$ is Haar random. 
$$
    e^{i\alpha H} = U D_\lambda U^\dagger, \quad \text{with} \quad
    D_\lambda = \text{diag}(e^{i\alpha\lambda_1}, \hdots, e^{i\alpha\lambda_d}).
$$
 As a result, the average in Eq. \eqref{eq:sm_avggue} can be decomposed into an average over Haar-distributed unitaries, together with an average over the eigenvalues encoded in $D_\lambda \otimes D^*_\lambda$, which must be done with respect to the GUE joint probability distribution.

\begin{equation}
    \overline{e^{i\alpha H}\otimes e^{-i\alpha H^*}}=\overline{(U\otimes U^*)\overline{(D_\lambda\otimes D^*_\lambda)}(U^\dagger \otimes U^T)} = \overline{(U\otimes U^*)\mathfrak{D}(U^\dagger \otimes U^T)}.
\end{equation}

After interchanging rows three and four, we identify the first box with Eq. \eqref{eq:sm_haar2copies}. Hence, in terms of the diagrammatic notation, we can write:
\begin{equation}
    \overline{e^{i\alpha H}\otimes e^{-i\alpha H*}}=\frac{1}{d^2-1}\Biggl(\Bigl( \boxtwotrA{\ \mathfrak{D}\ }-\frac{1}{d}\ \boxtwotrB{\ \mathfrak{D} \ }\Bigr)\ \ketub\braub \quad +\quad \Bigl( \boxtwotrB{\ \mathfrak{D} \ }-\frac{1}{d}\ \boxtwotrA{\ \mathfrak{D} \ }\Bigr)\ \idtwo\ \Biggr).
\end{equation}
The above expression can be further simplified taking into account
\begin{eqnarray}
        \boxtwotrA{\mathfrak{D}}\ &=& \ \braub\hspace{-0.5pt}\ketub = d\  \text{ since }\quad  D_\lambda D_\lambda ^T=\mathds{1}\, \nonumber \\
        \boxtwotrB{\mathfrak{D}}\ &=& \ \Biggl(\boxonetr{D_\lambda}\Biggr)^2 = \overline{\tr^2{D_\lambda}} = \overline{\sum_{m,n=1}^d e^{i\alpha (\lambda_m-\lambda_n})}. 
        \label{eq:sm_trboxes}
\end{eqnarray}
We recognize in the second row the definition of the \emph{spectral form factor} (SFF)~\cite{delcampoScramblingSpectralForm2017,liuSpectralFormFactors2018}: the Fourier transform of the two-point function. Observe that the strength of the noise $\alpha$ plays here the role of time. For the GUE we can eliminate the specific dependence on the labels of the eigenvalues:
\begin{align}
\label{fdalpha}
    f_d(\alpha)&:= f_d(\lambda_n,\lambda_m,\alpha) = \int \prod_{i=1}^d d\lambda_i  P_{\text{GUE}}(\lambda_1,\dots,\lambda_d)e^{i\alpha(\lambda_m-\lambda_n)}
\end{align}
where $P_{\text{GUE}}$ is the GUE probability density function. The explicit computation of $f_d(\alpha)$ can be found in Ref.~\cite{samosFidelityDecayError2025}. Now, it is enough to write the SFF as $\overline{\tr^2{D_\lambda}}=d(1+(d-1)f_d(\alpha)).$
Thus, the average Eq. \eqref{eq:sm_avggue} is
\begin{equation}
    \overline{e^{i\alpha H}\otimes e^{-i\alpha H^*}}=\frac{1-f_d(\alpha)}{d+1} \ \ketub\braub \ +\ \frac{df_d(\alpha)+1}{d+1}\  \idtwo\ .
    \label{eq:sm_avgguefinal}
\end{equation}
Next, we compute the product of the expression in Eq. \eqref{eq:sm_avgguefinal} ( previously adding two identities/ horizontal lines to the diagrams) with the average over four unitaries as described in Eq. \eqref{eq:sm_haar2copies}.
where we have used the overlaps computed before Eq.~\eqref{eq:sm_innerstates}. It is convenient to write the above expression in the state notation used throughout this appendix and omit the labeling of the gates, since the above result is valid for all dimensions:
\begin{eqnarray}
\label{eq:sm_u_nonorthogonal}
    \overline{\lp e^{i\alpha H}\otimes e^{-i\alpha H^*}\otimes\mathds{1}\otimes\mathds{1}\rp\lp R\otimes R^*\otimes R\otimes R^*\rp} 
    = \frac{1}{d^2-1}\biggl\{(1+\frac{f_d(\alpha)-1}{d(d+1)})\kket{\ptwo}\bbra{\ptwo}\nonumber \\
    +\frac{df_d(\alpha)+1}{d+1}\biggl(\kket{\pone}\bbra{\pone}-\frac{1}{d}\kket{\ptwo}\bbra{\pone}-\frac{1}{d}\kket{\pone}\bbra{\ptwo}\biggr)\biggr\}.
\end{eqnarray}

In terms of the spin basis $1/2$ Eq. \eqref{eq:sm_spinbasis} Eq. \eqref{eq:sm_u_nonorthogonal} becomes diagonal
\begin{align}
    &\overline{\lp e^{i\alpha H}\otimes e^{-i\alpha H^*}\otimes\mathds{1}\otimes\mathds{1}\rp\lp R\otimes R^*\otimes R\otimes R^*\rp}=\kket{\Up}\bbra{\Up}+\dfrac{df_d(\alpha)+1}{d+1}\kket{\Down}\bbra{\Down}.
    \label{eq:sm_avg_faulty_diag}
\end{align}

\subsubsection{\texorpdfstring{Calculation of \boldmath $\bbra{\Down}\overline{\tilde{\bm{\mathcal{V}}}}$ and $\bbra{\Up}\overline{\tilde{\bm{\mathcal{V}}}}$}{Calculation of Down and Up overlaps}}
\begin{align}
    \bbra{\Up}\overline{\tilde{\bm{\mathcal{V}}}}&=\bigotensor_{m=1}^{L/2}\Bigl((1-\kappa)\bbra{\Up_{2m-1,2m}}\Bigl(\kket{\Up_{2m-1,2m}}\bbra{\Up_{2m-1,2m}}+\dfrac{4f_4(\alpha)+1}{5}\kket{\Down_{2m-1,2m}}\bbra{\Down_{2m-1,2m}}\Bigr)\nonumber\\
    &+\kappa\overline{\sum^r_{\mu_{2m-1,2m}=1}\bbra{\Up_{2m-1,2m}}\bm{k}_{\mu_{2m-1,2m}}}\Bigr)=\bigotensor_{m=1}^{L/2}\Bigl((1-\kappa+\kappa)\bbra{\Up_{2m-1,2m}}=\bbra{\Up}=\bbra{0,1},
\end{align}
where in the last equality we have used again the magnon basis, which labels the states in terms of the number of sites with spin down.
\begin{align}
    \bbra{\Down}\overline{\tilde{\bm{\mathcal{V}}}}&=\frac{1}{d^2-1}\bigotensor_{m=1}^{L/2}\Biggl((1-\kappa)\left(\sqrt{15}\frac{4f_4(\alpha)+1}{5}\bbra{\Down_{2m-1,2m}}+\bbra{\Up_{2m-1,2m}}\right)\nonumber\\
    &\hspace{3cm}+\kappa\overline{\sum^r_{\mu_{2m-1,2m}=1}\left(\sqrt{15}\bbra{\Down_{2m-1,2m}}+\bbra{\Up_{2m-1,2m}}\right)\bm{k}_{\mu_{2m-1,2m}}}\Biggr)\nonumber\\
    &\hspace{2cm}-\frac{1}{d^2-1}\bigotensor_{m=1}^{L/2}\Biggl((1-\kappa)\bbra{\Up_{2m-1,2m}}+\kappa\overline{\sum^r_{\mu_{2m-1,2m}=1}\bbra{\Up_{2m-1,2m}}\bm{k}_{\mu_{2m-1,2m}}}\Biggr)
\end{align}

Since in this case we are averaging over both unitaries and Kraus noise, both of the options considered in the discussion around Eq.~\eqref{eq:sm_krauscase1} and Eq.~\eqref{eq:sm_krauscase2} hold. Hence, we can write 
\begin{equation}
\begin{split}
    &\overline{\sum^r_{\mu_{2m-1,2m}=1}\bbra{\Up_{2m-1,2m}}\bm{k}_{\mu_{2m-1,2m}}}=\bbra{\Up_{2m-1,2m}}\\
    &\overline{\sum^r_{\mu_{2m-1,2m}=1}\left(\sqrt{15}\bbra{\Down_{2m-1,2m}}+\bbra{\Up_{2m-1,2m}}\right)\bm{k}_{\mu_{2m-1,2m}}}=\bbra{\pone}\bm{k}_{\mu_{2m-1,2m}}=\bbra{\Up_{2m-1,2m}},
\end{split}
\end{equation}
and with it 
\begin{equation}
\begin{split}
    \bbra{\Down}\overline{\tilde{\bm{\mathcal{V}}}}&=\frac{1}{d^2-1}\Biggl(\bigotensor_{m=1}^{L/2}\left((1-\kappa)\sqrt{15}\frac{4f_4(\alpha)+1}{5}\bbra{\Down_{2m-1,2m}}+\bbra{\Up_{2m-1,2m}}\right)-\bbra{\Up}\Biggr)\\
    &=\frac{1}{\sqrt{d^2-1}}\sum_{m=1}^{L/2}(1-\kappa)^m (15)^{m/2}\left(\frac{4f_4(\alpha)+1}{5}\right)^m\sum_{i=1}^{\binom{L/2}{m}}\bbra{m,i},
\end{split}
\end{equation}
in complete analogy with Eq.~\eqref{eq:sm_overlapsV_magnonbasis}. Then it is easy to see that
\begin{equation}
\begin{split}
&\bbra{\Up}\overline{\bm{\tilde{\mathcal{V}}}}\kket{\Up}=1,\quad \bbra{\Up}\overline{\bm{\tilde{\mathcal{V}}}}\kket{\Down}=\bbra{\Down}\overline{\bm{\tilde{\mathcal{V}}}}\kket{\Up}=0\\
&\bbra{\Down}\overline{\bm{\tilde{\mathcal{V}}}}\kket{\Down}=\frac{1}{d^2-1}\sum_{m,n=1}^{L/2}(1-\kappa)^m 15^{\frac{m+n}{2}}\left(\frac{4f_4(\alpha)+1}{5}\right)^m\sum_{i,j=1}^{\binom{L/2}{m}\binom{L/2}{n}}\delta_{m,n}\delta_{i,j}\\
&\hspace{2cm}= \frac{\lp4+12f(\alpha)(1-\kappa)-3\kappa\rp^{L/2}-1}{d^2-1}\\
&\Bigl(\kket{\Up}\bbra{\Up}\overline{\bm{\tilde{\mathcal{V}}}}+\frac{\delta-1}{d^2-1}\kket{\Down}\bbra{\Down}\ \overline{\bm{\tilde{\mathcal{V}}}}\Bigr)^{T-1}= \kket{\Up}\bbra{\Up}+\left(\frac{\delta-1}{d^2-1}\right)^{T-1}\left(\bbra{\Down}\overline{\bm{\tilde{\mathcal{V}}}}\kket{\Down}\right)^{T-2}\kket{\Down}\bbra{\Down}\overline{\bm{\tilde{\mathcal{V}}}}\ .
\end{split}
\end{equation}
Plugging them into Eq.\eqref{eq_sm_3partsfid2} yields the final result:
\begin{equation}
        \overline{\mathcal{F}_\text{simp}}=\frac{1}{d}+\left(\frac{\delta-1}{d^2-1}\right)^{T}\left(\bbra{\Down}\overline{\bm{\tilde{\mathcal{V}}}}\kket{\Down}\right)^{T}(1-\frac{1}{d}),\quad \delta=\overline{\tr^2{\tilde{\Pi}\Pi^T}}
\end{equation}
\bibliographystyle{quantum}
\setcitestyle{maxnames=5}
\bibliography{all_zotero}

\end{document}